\documentclass[%
12pt,T
 oneocolumn,
 nofootinbib,
 amsmath,amssymb,
 aps,
floatfix,
unsortedaddress
]{revtex4-2}

\usepackage{graphicx,color}
\usepackage{graphicx}
\usepackage{epstopdf}
\usepackage{subfig}
\usepackage{dcolumn}
\usepackage{bm}
\usepackage{graphicx,color}
\usepackage{float}
\usepackage{multirow}
\usepackage{amsmath}
\usepackage{mathtools}
\usepackage{cancel}
\usepackage{subcaption}
\usepackage{subfig}
\usepackage[pdfpagelabels, pdfencoding=auto, psdextra]{hyperref}
\hypersetup{%
 pdfsubject=Paper,
 pdfkeywords={nuclear physics} {Pionless EFT} {Two-Nucleon} {Lattice} {L\"usher formula},
 unicode = true,
 breaklinks = true,
 colorlinks = true,
 linkcolor = blue,
 citecolor = blue,
 menucolor = blue,
 citecolor = blue,
 urlcolor = blue
}

\graphicspath{{./}}
\begin{document}

\title{	Examination of the $c\bar{c}+n+^{10}$Be bound-state problem within three cluster models based on QCD charmonium-nucleon interactions }
\author{Faisal Etminan}
 \email{fetminan@birjand.ac.ir}
\affiliation{
 Department of Physics, Faculty of Sciences, University of Birjand, Birjand 97175-615, Iran
}%
\affiliation{ Interdisciplinary Theoretical and Mathematical Sciences Program (iTHEMS), RIKEN, Wako 351-0198, Japan}

\date{\today}%
\begin{abstract}
	The possible existence of a bound state in the charmonium--neutron--$^{10}$Be system is investigated within a three-cluster model in which the system is treated as $c\bar{c}+n+^{10}$Be. The three-body problem is solved using a hyperspherical-harmonics expansion within the coupled Faddeev-equation framework. The neutron--$^{10}$Be interaction is described by an $l$-dependent Woods--Saxon potential that provides a satisfactory description of the low-energy properties of $^{11}$Be.
	 The effective $^{10}$Be--$c\bar{c}$ interactions are constructed by folding the $S$-wave charmonium--nucleon potentials obtained from the HAL QCD Collaboration at nearly physical pion mass with the $^{10}$Be matter density. 	%
	Bound three-body states are obtained for all three spin-dependent charmonium--nucleon interactions considered. The central three-body binding energies are $3.297(493)$ MeV for $J/\psi N(^4S_{3/2})$, $4.650(632)$ MeV for $J/\psi N(^2S_{1/2})$, and $2.009(325)$ MeV for $\eta_cN(^2S_{1/2})$, with corresponding nuclear matter rms radii of $2.542(24)$, $2.474(17)$, and $2.680(42)$ fm, respectively. 
		These results demonstrate that the attractive charmonium--nucleon interactions obtained from lattice QCD can generate bound $c\bar{c}+n+^{10}$Be configurations within the present three-body framework, while the predicted binding energies retain a substantial sensitivity to the spin dependence of the underlying interaction.
\end{abstract}


\maketitle
\section{Introduction} \label{sec:intro}
The possible existence of bound states formed by charmonium and nuclei has attracted considerable theoretical and experimental interest because such states would provide a unique probe of the interaction between compact heavy-quark systems and nuclear matter. In particular, the observation of a $J/\psi$--nuclear bound state would provide evidence for an attractive interaction between charmonium and nuclear matter and could offer insight into the role of gluonic degrees of freedom in hadronic systems~\cite{BrodskyPRL1990,VOLOSHIN2008455}. Such states are also expected to exhibit relatively narrow structures, making them attractive targets for future searches at facilities such as Jefferson Lab, J-PARC, and FAIR~\cite{KreinPRC2011}.

Unlike ordinary hadron--nucleus interactions, the interaction between charmonium and a nucleon is not naturally described in terms of conventional light-meson exchange. Because the $c\bar c$ state contains no light valence quarks, the coupling to ordinary light-quark exchange mechanisms is strongly suppressed by the OZI rule. The interaction is therefore expected to be governed predominantly by gluonic dynamics, often described in terms of QCD van der Waals forces or multipole gluon exchange. Early theoretical studies suggested that such interactions could be sufficiently attractive to produce charmonium--nuclear bound states~\cite{BrodskyPRL1990,WassonPRL1991}, while subsequent approaches based on QCD sum rules, effective field theories, quark models, and phenomenological potentials have generally predicted attractive charmonium--nucleus interactions, although with substantial differences in the resulting binding energies~\cite{KIM2001517,PhysRevC.82.045207,PhysRevD.71.076005,PhysRevC.63.044906,PhysRevC.75.064903,PhysRevC.75.064907,KreinPRC2011,ETMINAN2025PLB139564,etminan2026SR}.

Along similar lines, the inclusion of nuclear density distributions allowed estimates of $\eta_c$–nuclear bound state energies as presented in Ref.~\cite{PhysRevD.58.034012}. Energies of $\eta_c$–nucleus bound states for a variety of nuclei were also evaluated in Ref.~\cite{COBOSMARTINEZ2020135882}, where key inputs—including in‑medium modified masses of $D$ and $D^{*}$ mesons and nuclear density profiles—were derived using the quark–meson coupling model.

A major recent development has been the direct determination of low-energy charmonium--nucleon interactions from lattice QCD using the HAL QCD method~\cite{ishii2012, aoki2013,Iritani2019prb, sasaki2020, etminan2024prd}. In particular, the HAL QCD Collaboration has recently obtained $S$-wave interactions for the $J/\psi N(^4S_{3/2})$, $J/\psi N(^2S_{1/2})$, and $\eta_cN(^2S_{1/2})$ channels at nearly physical pion mass~\cite{LyuPLB2025}. These interactions are attractive over the relevant distance range and exhibit a common long-range behavior, while their short-range strengths retain a non-negligible dependence on the spin channel. The availability of such QCD-based interactions provides an opportunity to revisit charmonium--nuclear binding using nuclear models in which the microscopic input is constrained directly by lattice QCD rather than by phenomenological charmonium--nucleus potentials.

Several studies have investigated the formation of charmonium--nuclear bound states using density-based and cluster approaches. Lattice QCD calculations at an $SU(3)$ flavor-symmetric point with an unphysically large pion mass have also reported bound quarkonium--nucleus systems~\cite{Beane2015}. More recently, cluster calculations of $c\bar c+\alpha+\alpha$ systems have demonstrated that few-body methods based on hyperspherical harmonics can provide a useful framework for studying the structure and binding of charmonium-containing nuclear systems~\cite{EtminanPRCalphaAlphaccbar2025}. These developments motivate the study of more neutron-rich and spatially extended nuclear configurations in which the interplay between a weakly bound neutron and an attractive charmonium--core interaction can be investigated explicitly.

In the present work, the hypothetical three-body system $c\bar c+n+{}^{10}\mathrm{Be}$ is considered, with $^{10}$Be treated as an inert $0^+$ core. This system is particularly interesting because the $n+{}^{10}$Be subsystem forms the weakly bound halo nucleus $^{11}$Be. An $l$-dependent Woods--Saxon neutron--core interaction based on the interaction of Thompson \textit{et al.}~\cite{ThompsonPRC1996}, is employed. The original interaction is modified only through its Woods--Saxon strengths in order to remove deeply bound Pauli-forbidden states while preserving the geometry and reproducing the experimentally known low-energy properties of $^{11}$Be. The resulting interaction is independently tested against the neutron separation energy, neutron--core rms separation, matter and charge radii, asymptotic wave number, and asymptotic normalization coefficient. This validation is important because the extended spatial structure of the $^{11}$Be subsystem can substantially influence the properties of the three-body state.

For the $^{10}$Be--$c\bar c$ interaction, the recently determined HAL QCD charmonium--nucleon potentials~\cite{LyuPLB2025} is employed. These interactions are folded with the ground-state matter density of $^{10}$Be using the single-folding approximation. The resulting numerical potentials are represented by Woods--Saxon plus Gaussian forms, which reproduce the folding potentials with deviations much smaller than the statistical uncertainties inherited from the HAL QCD input. The statistical uncertainties of the fitted effective interactions are subsequently propagated to the calculated three-body binding energies and matter radii.

The three-body problem is solved using a hyperspherical-harmonics expansion within the coupled Faddeev-equation framework~\cite{face}. Particular attention is paid to the numerical convergence with respect to the maximum hypermomentum $K_{\max}$ and to the composition of the resulting three-body wave functions. In addition to the central binding energies and matter radii, the probabilities of the dominant hyperspherical configurations are calculated in order to characterize the spatial and angular structure of the predicted bound states.

The main objective of this work is therefore to determine whether the QCD-based charmonium--nucleon interactions obtained from lattice QCD can generate a bound $c\bar c+n+{}^{10}$Be system when combined with a realistic description of the weakly bound $n+{}^{10}$Be subsystem. The dependence of the three-body observables on the spin structure of the charmonium--nucleon interaction is also examined, while the propagation of the corresponding statistical uncertainties is quantified. This provides a direct connection between the microscopic QCD input and the predicted properties of a possible charmonium--nuclear bound state.

The paper is organized as follows.  The three-body hyperspherical-harmonics formalism is summarized in Sec.~\ref{sec:Three-body hyperspherical}. Section~\ref{sec:Two-body-potentials} describes the two-body interactions and the construction and validation of the effective $^{10}$Be--$c\bar c$ potentials. The numerical results, convergence properties, uncertainty propagation, and dominant hyperspherical components are presented in Sec.~\ref{Three-cluster systems}. Finally, the main conclusions and implications of the results are summarized in Sec.~\ref{sec:Summary-and-conclusions}.

\section{ Hyperspherical harmonics formalism  }
\label{sec:Three-body hyperspherical}
This section provides a concise overview of the formalism employed, drawing upon extensive prior studies~\cite{Zhukov93, raynal1970, face,etminan2024prc}. The aim is to define the key quantities and notation essential for subsequent sections.

\begin{figure*}[hbt!]
	\centering
	\includegraphics[scale=0.20]{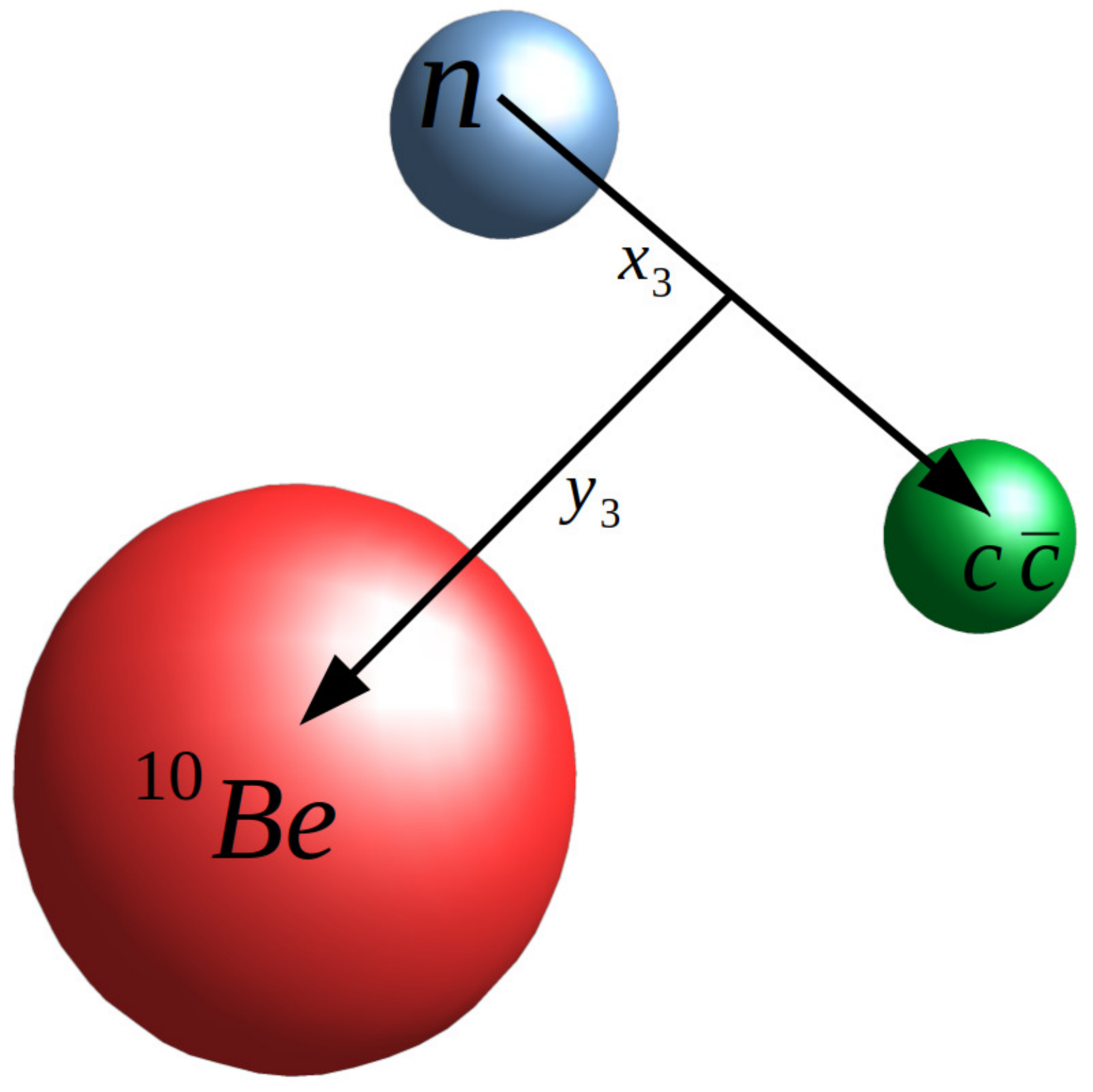}
	\caption{The Jacobi-T set, connects $c\bar{c}$ and $ n $ particles by coordinate $\vec{x}$, is used to describe $ c\bar{c}+n+^{10}$Be system. 
	}
	\label{fig:T_jacobi}
\end{figure*}

The model under consideration is a three-body system composed of a \(c\bar{c}\) pair, an \(n\) particle, and a \(^{10}\textrm{Be}\) nucleus, designated as the core. These components are indexed as \(i = 1, 2, 3\), with associated masses \(m_i\), position vectors \(r_i\), and momenta \(p_i\). Interactions between these particles are mediated by pairwise potentials \(V_{ij}\).
The system is described using Jacobi coordinates. These coordinates, apart from mass factors, represent the relative position between two particles (denoted by \(\boldsymbol{x}\)) and the position vector between their combined center of mass and the third particle (denoted by \(\boldsymbol{y}\)). An illustration of these coordinates can be found in Fig.~\ref{fig:T_jacobi}. The mathematical expressions are as follows:
\begin{equation}
	\vec{x}_{i}=\sqrt{\frac{m_{j}m_{k}}{m\left(m_{j}+m_{k}\right)}}\ \vec{r}_{j}-\vec{r}_{k},\quad\vec{y}_{i}=\sqrt{\frac{m_{i}(m_{j}+m_{k})}{m\left(m_{i}+m_{j}+m_{k}\right)}}\left(\vec{r}_{i}-\frac{m_{j}\vec{r}_{j}+m_{k}\vec{r}_{k}}{m_{j}+m_{k}}\right),
\end{equation}
here, $m = m_N = 938.9$ MeV serves as a normalization mass.

The precise definitions of these coordinates, along with the three corresponding sets of hyperspherical coordinates \(\{\rho, \theta, \Omega_x, \Omega_y\}\), are detailed in Refs.~\cite{Zhukov93, raynal1970, face}. In this context, \(\rho^2 = x^2 + y^2\) represents the generalized radial coordinate, while the hyperangle \(\theta = \arctan(x/y)\), ranging from $0$ to $\pi/2$, quantifies the relative magnitudes of \(\boldsymbol{x}\) and \(\boldsymbol{y}\). The angles \(\Omega_{x}\) and \(\Omega_{y}\) define the orientations of \(\boldsymbol{x}\) and \(\boldsymbol{y}\), respectively. For conciseness, all angular dependencies are collectively represented by the set $\varphi \equiv (\theta, \Omega_x, \Omega_y)$.

In a three-body model involving three particles or clusters, such as \(c\bar{c}+n+^{10}\textrm{Be}\), the total three-body wave function \(\Psi\) is expressed as $ \Psi=\sum_{i=1}^{3}\psi^{\left(i\right)} $. 
Each component, \(\psi^{(i)}\), is a function of a distinct set of Jacobi coordinates, one of which is depicted in Fig.~\ref{fig:T_jacobi}. These components adhere to the coupled Faddeev equations:
\begin{equation}
	\left(T_{i}+h-E\right)\psi^{\left(i\right)}+V_{i}\left(\psi^{\left(i\right)}+\psi^{\left(j\right)}+\psi^{\left(k\right)}\right)=0,
	\label{eq:faddeev_coupled-eq}
\end{equation}
where $T$ is the kinetic energy operator, $E$ denotes the total energy, $h=\sum_{i}h_{i}$ represents the sum of the intrinsic Hamiltonians of each particle $h_{i}$, and $V_{jk}(r_{jk})$ is the interaction potential between particles $j$ and $k$. The indices $i, j, k$ are subject to cyclic permutation of (1, 2, 3).

Within the context of a three-body \(c\bar{c}+n+\text{core}\) model, the total wave function is assumed to have a product form:
\begin{equation}
	\psi^{(i)}\left(x_{i},y_{i},\zeta_{c}\right)=\phi_{c}\left(\zeta_{c}\right)\psi_{c}\left(x_{i},y_{i}\right),
\end{equation}
here, \(\phi_c(\zeta_c)\) represents the intrinsic wave function of the core, and \(\psi_c\) encompasses the radial, angular, and spin degrees of freedom of the two particles relative to the core. The core's eigenstates \(\phi_c\) and eigenvalues \(\varepsilon_c\) are determined by the core's intrinsic Hamiltonian, $\hat{h}_c\phi_c(\zeta_c)=\varepsilon_c\phi_c(\zeta_c)$. The internal coordinates of the core, \(\zeta_c\), must be included alongside the Jacobi coordinates to fully define the quantum state of the system.

For states with a fixed total angular momentum \(j\), the wave function is expanded, for each hyperradius \(\rho\), as
\begin{equation}
\psi^{j}\left(\rho,\varphi\right)=\rho^{-5/2}\sum_{\beta}\mathcal{R}^{\beta j}\left(\rho\right)\,\mathcal{Y}^{\beta j}\left(\varphi\right),
	\label{eq:hf}
\end{equation}
where the channel label \(\beta\) is defined as
$ \beta \equiv \left\{ K, l_x, l_y, l, S_x, j_{ab}\right\}$.
The quantity $K=l_{x}+l_{y}+2n$ $\left(n=0,1,2,...\right)$ is the hypermomentum. The orbital angular momenta \(l_x\) and \(l_y\) correspond to the Jacobi coordinates \(\boldsymbol{x}\) and \(\boldsymbol{y}\), respectively. The total orbital angular momentum is
$ \mathbf{l} = \mathbf{l_x} + \mathbf{l_y} $. Furthermore, \(S_x\) denotes the spin of the two particles coupled through the coordinate \(\boldsymbol{x}\). The intermediate angular momentum \(j_{ab}\) arises from the coupling $ \mathbf{j}_{ab} = \mathbf{l} + \mathbf{S_x} $. If \(\mathbf{I}\) is the spin of the third particle (assumed fixed), then the total angular momentum \(j\) is obtained from $ \mathbf{j} = \mathbf{j}_{ab} + \mathbf{I} $. 

In the present calculations, the $^{10}$Be core is treated within a two-state coupled-channel approximation that includes its $0_1^+$ ground state and first excited $2_1^+$ state. Consequently, the core spin can take the values $I=0$ and $I=2$, and the total angular momentum of the three-body system is obtained by coupling the angular momentum of the $c\bar{c}+n$ subsystem, $j_{ab}$, with the core spin according to
$ \mathbf{j}=\mathbf{j}_{ab}+\mathbf{I} $.
For the $0_1^+$ core configuration, $I=0$ and hence $j=j_{ab}$. For the $2_1^+$ configuration, the allowed total angular momenta are obtained from the corresponding angular-momentum coupling. In the following discussion and tables, the quantum number $j$ denotes the total angular momentum of the complete $c\bar{c}+n+{}^{10}$Be system.

The hyperangular functions $\mathcal{Y}^{\beta j}\left(\varphi\right)$ are constructed from hyperspherical harmonics, which are eigenstates of the hypermomentum operator \(\hat{K}\)~\cite{Zhukov93,Casal2020,raynal1970}.
The hyperradial functions $\mathcal{R}^{\beta j}\left(\rho\right)$ describe the hyperradial dependence. To facilitate the solution of the resulting coupled differential equations~\cite{face}, they are expanded in a basis of square-integrable functions:
\begin{equation}
\mathcal{R}^{\beta j}\left(\rho\right)=\sum_{i=0}^{i_{\max}}C_{i}^{\beta j}\,U_{i}^{\beta}\left(\rho\right).
\end{equation}
Here, \(i\) denotes the hyperradial excitation index, so that \((i_{\max}+1)\) is the number of hyperradial basis functions included in each channel. The coefficients $C_{i}^{\beta j}$ are the (diagonalization) amplitudes associated with the chosen basis.

The normalization of the three-body wave function satisfies $\langle \Psi^{j} | \Psi^{j} \rangle = 1$.
Accordingly, the contribution (or weight) of a given hyperspherical channel $\beta$ to the total wave function is defined as
\begin{equation}
P^{\beta K}=\int_{0}^{\infty}\left|\mathcal{R}^{\beta K}\left(\rho\right)\right|^{2}d\rho,
\end{equation}
which represents the probability of finding the system in the channel $\beta$. Owing to the orthonormality of the hyperspherical basis, the channel probabilities satisfy the normalization condition $\sum_{\beta K}P^{\beta K}=1$.
Similarly, the total contribution of a given hypermomentum $K$ is obtained by summing over all channels having the same value of $K$,
$P^{K}=\sum_{\beta}P^{\beta K}$,
where the summation extends over all channels characterized by the hypermomentum $K$. The quantities $P^{\beta}$ and $P^{K}$ provide a quantitative measure of the importance of the individual hyperspherical configurations and are useful for assessing the convergence of the hyperspherical expansion.

The channel probabilities provide a quantitative measure of the relative importance of the different hyperspherical configurations in the total wave function and constitute a useful indicator of the convergence of the hyperspherical expansion.

The hyperradial functions in  Eq.~\eqref{eq:hf} are solutions to the following set of coupled differential equations
\begin{equation}
	\left(-\frac{\hbar^{2}}{2m}\left(\frac{d^{2}}{d\rho^{2}}-\frac{(K+3/2)(K+5/2)}{\rho^{2}}\right)-E\right)\mathcal{R}^{\beta j}\left(\rho\right) +\sum_{\beta'}V_{\beta'\beta}^{j}(\rho)\mathcal{R}^{\beta' j}\left(\rho\right)=0,\label{eq:Faddeev_{c}oupled}
\end{equation}
In this equation, $K$ defines an effective three-body barrier, and the term \(V_{\beta' \beta}^{j}(\rho)\) represents the coupling potentials. These potentials, which arise from the two-body interactions \(V_{ij}\) described in Section~\ref{sec:Two-body-potentials}, are defined as:
\begin{equation}
V_{\beta'\beta}^{j}\left(\rho\right)=\left\langle \mathcal{Y}^{\beta j}\left(\varphi\right)\left|V_{12}+V_{13}+V_{23}\right|\mathcal{Y}^{\beta'j}\left(\varphi\right)\right\rangle. \label{eq:V_ij}
\end{equation}

The matter radius of an \(A\)-nucleon system is evaluated according to
\begin{equation}
r_{\text{mat}}=\sqrt{\frac{1}{A}\left[\left(\sum_{q=1}^{3}A_{q}\left\langle r_{A_{q}}^{2}\right\rangle \right)+\left\langle \rho^{2}\right\rangle \right]}, \label{eq:r_mat}
\end{equation}
where $A_{q}$ and $\left\langle r_{A_{q}}^{2}\right\rangle $ denote the mass number and the intrinsic mean-square radius of the q-th cluster, respectively.
The term $\left\langle \rho^{2}\right\rangle$  represents the contribution arising from the relative motion between the clusters and is calculated from the three-body wave function for the state under consideration. 
The matter radii of the \( c\bar{c}+n+^{10}\)Be ground state were obtained using the experimental rms matter radius of (\(^{10}\)Be) = 2.30 fm~\cite{TANIHATA1988592}. A rms radius of 0.8 fm was adopted for the neutron. In addition, a strong-interaction radius of 0.25 fm was assumed for the \( c\bar{c} \) pair~\cite{POVH1990653}. The resulting matter radii of the calculated \( c\bar{c}+n+^{10}\)Be bound states are summarized in  Table~\ref{tab:Charm-alpha-para}.

\section{Two-cluster subsystems }
\label{sec:Two-body-potentials}
\subsection{$ c\bar{c} $-N interactions}\label{subsec:charm-N_pot}
A realistic lattice QCD simulation of the $S$‑wave $c\bar{c}\textrm{-}N$ potentials has been presented by the HAL QCD Collaboration evaluated near the physical pion mass $m_{\pi} = 146.4(4)$ MeV at the imaginary‑time slice $t/a = 14$, with lattice spacing $a = 0.0846$ fm~\cite{LyuPLB2025}. The spectroscopic notation $^{2s+1}L_{J}$ is used, where $s$ denotes the total spin, and $L$ and $J$ denote the orbital and total angular momenta, respectively.

\begin{figure*}[hbt!]
	\centering
	\includegraphics[scale=1.0]{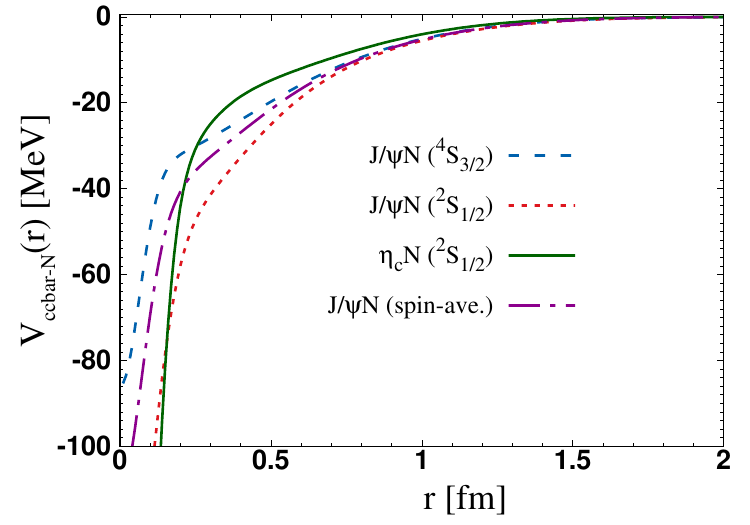}
	\caption{
		The \( S \)-wave \( c\bar{c}\textrm{-}N \) potentials given in Eq.~\eqref{eq:NCharm_pot} are plotted as functions of the separation between \( N \) and \( c\bar{c} \) at an imaginary‑time slice \( t/a = 14 \). The parametrization follows Ref.~\cite{LyuPLB2025} and employs the constants listed in Table~\ref{tab:Charm-N-para}. The spin‑\(3/2\) \( J/\psi N \) potential is represented by the dashed blue line, the spin‑\(1/2\) \( J/\psi N \) potential by the dotted red line, the spin‑\(1/2\) \( \eta_{c} N \) potential by the solid green line, and the spin‑averaged \( J/\psi N \) potential (Eq.~\eqref{eq:NJpsi-spin-ave}) by the dash‑dotted purple line. 
		\label{fig:CharmN} }
\end{figure*}

The interactions, which are extracted from spacetime correlation functions of the charmonium–nucleon system using the HAL QCD method, are found to be attractive at all distances and to exhibit a long‑range tail consistent with a two‑pion‑exchange potential. It is reported in Ref.~\cite{LyuPLB2025} that the $J/\psi N$ potential is slightly stronger than the $\eta_c N$ potential at short distances, as illustrated in Fig.~\ref{fig:CharmN}. In contrast, the $J/\psi N$ ($^{4}S_{3/2}$), $J/\psi N$ ($^{2}S_{1/2}$), and $\eta_c N$ ($^{2}S_{1/2}$) potentials display similar behavior at large distances and become nearly degenerate, indicating that a common underlying mechanism governs these interactions at long range.

In Ref.~\cite{LyuPLB2025}, for the imaginary‑time slice $t/a = 14$, an uncorrelated fit is performed in the range $0 \le r \le 1.8$ fm using phenomenological three‑range Gaussian functions,
\begin{equation}
	V_{c\bar{c}\textrm{-}N}\left(r\right)=-\sum_{n=1}^{3}a_{n}e^{-\left(r/b_{n}\right)^{2}}. \label{eq:NCharm_pot}
\end{equation} 
 The fitted parameters are taken directly from Ref.~\cite{LyuPLB2025} and are listed in Table~\ref{tab:Charm-N-para}. The quality of the fit is illustrated in Fig.~\ref{fig:CharmN}.

Because current measurements of two‑body correlation functions and scattering parameters are restricted to spin‑averaged quantities, the spin‑averaged \( J/\psi N \) potential is defined as  
\begin{equation}
	J/\psi N^{\textrm{spin-ave}}=\frac{2}{3}J/\psi N\left(^{4}S_{3/2}\right)+\frac{1}{3}J/\psi N\left(^{2}S_{1/2}\right). \label{eq:NJpsi-spin-ave}
\end{equation}
This definition enables the evaluation of the interaction independently of detailed spin correlations.  
Figure~\ref{fig:CharmN} reveals qualitative distinctions among the low‑energy spin‑dependent \( c\bar{c}\textrm{-}N \) potentials. Examining the influence of these variations is particularly valuable for understanding their effect on the binding energy of the \( c\bar{c}+n+^{10}\textrm{Be} \) ground state within the three‑body system. 

\begin{table}[hbt!]
	\centering
	\caption{
		The fitting parameters included in Eq.~\eqref{eq:NCharm_pot}, together with their statistical errors (shown in parentheses), correspond to the time slice \( t/a = 14 \). The parameters $a_n$ (MeV) and $b_n$ (fm), as well as the scattering length $a_0$ (fm) and effective range $r_e$ (fm), are also listed. For the latter two quantities, the first and second parentheses indicate the statistical and systematic uncertainties, respectively. All parameter values are taken directly from Ref.~\cite{LyuPLB2025}.
			\label{tab:Charm-N-para}}	
	\begin{tabular}{ccccccccc}
		\hline
		\hline 
		& $a_{1}$ & $b_{1}$& $a_{2}$&$b_{2}$&$a_{3}$&$b_{3}$&$ a_0$&$ r_{e}$\\
		\hline
		$ J/\psi N\left(^{4}S_{3/2}\right) $& $ 51(1)$  & $ 0.09(1) $ & $ 13(6)$ & $ 0.49(7)$ & $22(5)$& $0.82(6)$&$0.30(2)_{+0}^{-2}$&$3.25(12)_{+6}^{-9}$ \\	
		$ J/\psi N\left(^{2}S_{1/2}\right) $& $ 101(1) $& $ 0.13(1) $ & $33(6)$  & $ 0.44(5)$ & $23(8)$& $0.83(9)$&$0.38(4)_{+0}^{-3}$&$2.66(21)_{+0}^{-10}$ \\
		$\eta_{c} N\left(^{2}S_{1/2}\right)$& $ 264(14) $& $ 0.11(1) $ & $28(13)$ & $0.24(6)$  & $22(2)$& $0.77(3)$&$0.21(2)_{+0}^{-1}$&$3.65(19)_{+0}^{-6}$ \\
		\hline
		\hline 	
	\end{tabular}
\end{table}
As explained in the Introduction, the low‑energy interaction between a nucleon and a heavy quarkonium (\( Q\bar{Q} \)) is thought to arise mainly through multiple‑gluon exchanges. Because color confinement prevents gluons from propagating over long distances, the effective degrees of freedom in that regime are color‑neutral states coupled to gluons. Among these, the two‑pion state is expected to play a dominant role~\cite{Brambilla2016}. Evidence presented by Lyu \textit{et al.} in Ref.~\cite{LyuPLB2025} confirms that their lattice QCD results are consistent with a long‑range \( c\bar{c}\textrm{-}N \) interaction driven by a two‑pion‑exchange mechanism for \( Q\bar{Q}\textrm{-}N \) systems~\cite{PhysRevD.98.014029}.

\subsection{Construction of $ ^{10}\textrm{Be}\textrm{-}c\bar{c} $ potential via the single-folding method} \label{sec:folding-Model}
Since the $^{10}\mathrm{Be}$ nucleus is strongly bound and its internal structure is expected to be only weakly modified by the interaction with charmonium, the effective $^{10}\mathrm{Be}\textrm{-}c\bar{c}$ interaction is constructed within the single-folding approximation. In this framework, the charmonium interacts independently with the individual nucleons inside the target nucleus, while many-body corrections beyond the impulse approximation are neglected. The resulting effective interaction is obtained by folding the HAL QCD charmonium--nucleon interaction with the ground-state matter density of the $^{10}\mathrm{Be}$,
\begin{equation}
	U_{^{10}\textrm{Be}\textrm{-}c\bar{c}}(R)
	=
	\int
	\rho(\mathbf{r})
	V_{c\bar{c}\textrm{-}N}
	\!\left(
	|\mathbf{R}-\mathbf{r}|
	\right)
	d^{3}r,
	\label{eq:folding}
\end{equation}
where $R$ denotes the distance between the center of mass of the $^{10}\mathrm{Be}$ nucleus and the charmonium~\cite{Satchler1979, Etminan:2019gds}. The interaction $V_{c\bar{c}\textrm{-}N}$ is taken directly from the HAL QCD calculation and is represented by the three-range Gaussian parametrization introduced in subsection~\ref{subsec:charm-N_pot}.

Assuming spherical symmetry, Eq.~(\ref{eq:folding}) reduces to
\begin{equation}
	U(R)
	=
	2\pi
	\int_{0}^{\infty}
	r^{2}\rho(r)
	\int_{-1}^{1}
	V_{c\bar{c}\textrm{-}N}
	\!\left(
	\sqrt{R^{2}+r^{2}-2Rr\mu}
	\right)
	d\mu\,dr,
	\label{eq:folding_radial}
\end{equation}
which is evaluated numerically using adaptive Gauss--Kronrod quadrature. The numerical convergence of the folding integral was verified by varying the radial integration limit and the numerical tolerances until the resulting potential became stable within numerical precision.

The matter density of the $^{10}\textrm{Be}$ nucleus is taken from Ref.~\cite{Lenz1993} and parameterized by a two-parameter Fermi distribution,
\begin{equation}
	\rho(r)
	=
	\frac{\rho_{0}}
	{1+\exp\!\left[(r-R_D)/a_D\right]},
\end{equation}
where
$ R_D =	1.12A^{1/3} - 0.86A^{-1/3}\ {\rm fm} $, with $A=10$, diffuseness $a_D=0.54$ fm, and normalization
$ 4\pi \int_{0}^{\infty} \rho(r)r^{2}dr = A $, which yields $ \rho_0=0.171~\mathrm{fm^{-3}} $.

At sufficiently large separations between \(^{10}\textrm{Be}\) and \(c\bar{c}\), the system is described effectively as a \(^{10}\textrm{Be}+c\bar{c}\) cluster. In contrast, in and near the \(^{10}\textrm{Be}\) region, all possible cluster configurations should, in principle, contribute. Since no bound states are generated by the \(c\bar{c}\) meson in any two‑body subsystem, the \(^{10}\textrm{Be}+c\bar{c}\) clusterization is assumed to dominate. Under this assumption, the folding potential is regarded as an appropriate approximation to the \(^{10}\textrm{Be}\textrm{-}c\bar{c}\) interaction. As mentioned, this constitutes a simplifying assumption: experimentally, the \(^{10}\textrm{Be}\textrm{-}c\bar{c}\) interaction is expected to be effectively spin‑averaged (with respect to spin‑dependent nucleon–\(J/\psi\) interactions), since the ten nucleons in \(^{10}\textrm{Be}\) are not spin‑polarized.

The statistical uncertainties of the HAL QCD interaction are propagated to the folded potential through a Monte Carlo sampling procedure~\cite{JCGM101}. Since the covariance matrix of the Gaussian fit parameters is not available in Ref.~\cite{LyuPLB2025}, each Gaussian strength and range parameter is sampled independently according to a normal distribution using the quoted statistical uncertainty as its standard deviation. For every Monte Carlo realization, the folding integral is recalculated, generating an ensemble of effective $^{10}\textrm{Be}\textrm{-}c\bar{c}$ potentials. The mean interaction together with the corresponding $68\%$ and $95\%$ confidence intervals are then extracted at each radial distance.

To characterize the folded interaction, two integral quantities are additionally calculated, namely the volume integral,
\begin{equation}
	J_V = 	4\pi \int_{0}^{\infty} U(R)R^{2}dR,
\end{equation}
and the rms radius of the potential,
\begin{equation}
	R_{\rm rms}
	=
	\left[
	\frac{
		\displaystyle
		\int_{0}^{\infty}
		R^{4}U(R)dR
	}{
		\displaystyle
		\int_{0}^{\infty}
		R^{2}U(R)dR
	}
	\right]^{1/2},
\end{equation}
which provide global measures of the interaction strength and its spatial extension.

For practical implementation in the subsequent three-body calculations, each numerically obtained folding potential is further parameterized by a Woods--Saxon potential supplemented by a Gaussian correction,
\begin{equation}
	U(R) =	- \frac{V_0} {1+\exp\!\left[(R-R_0)/a\right]}
	-
	V_G
	\exp\!\left[
	-\left(
	\frac{R}{b_G}
	\right)^2
	\right],
	\label{eq:WSG}
\end{equation}
where the parameters ${V_0,R_0,a,V_G,b_G}$ are determined independently for each of the four channels through a nonlinear least-squares fit to the corresponding numerical folding potential.

\begin{figure*}[t]
	\centering
	
	\subfloat[$J/\psi N(^4S_{3/2})$]{
		\includegraphics[width=0.45\textwidth]
		{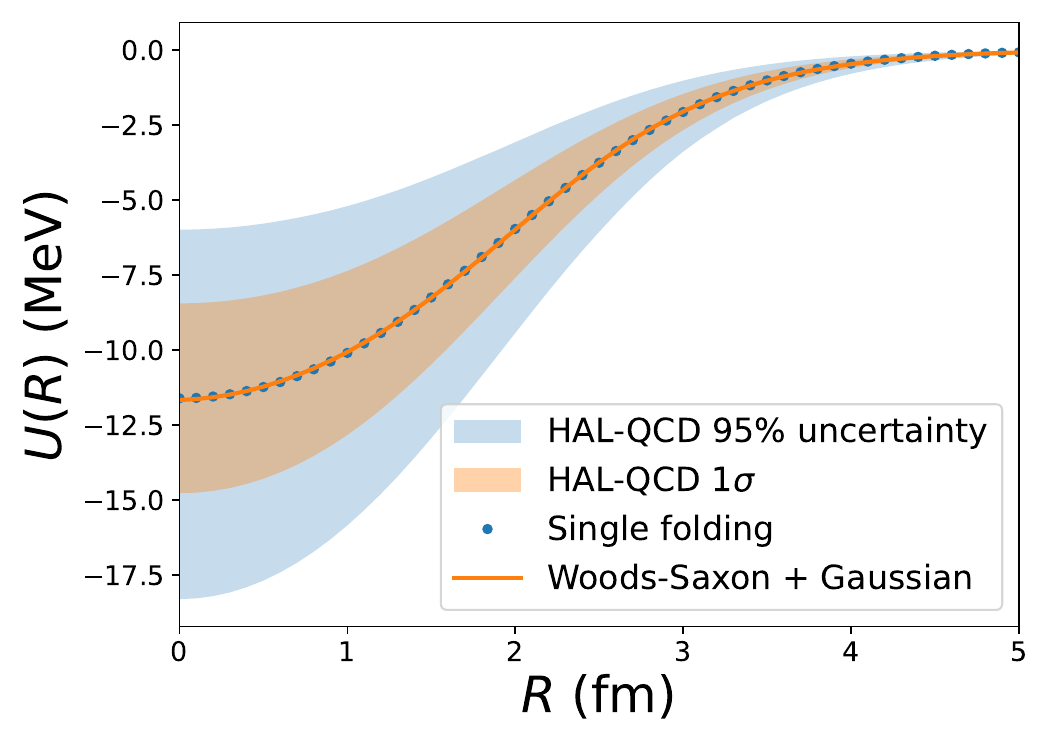}
	}
	\hfill
	\subfloat[$J/\psi N(^2S_{1/2})$]{
		\includegraphics[width=0.45\textwidth]
		{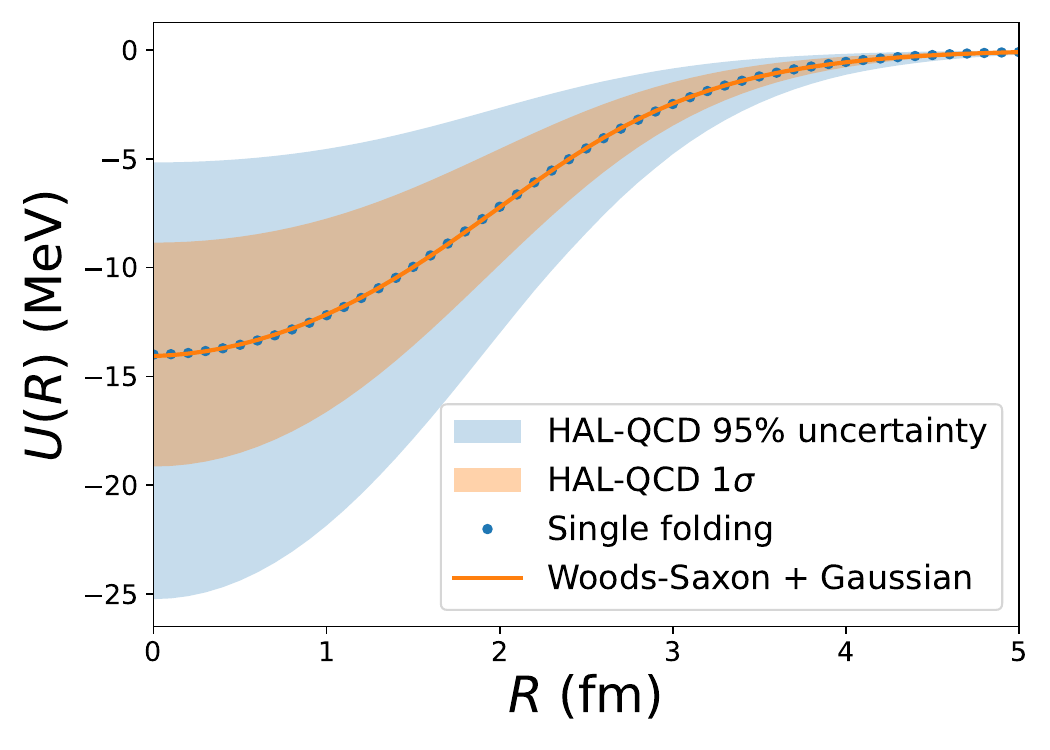}
	}
	\hfill
	\subfloat[$\eta_cN(^2S_{1/2})$]{
		\includegraphics[width=0.45\textwidth]
		{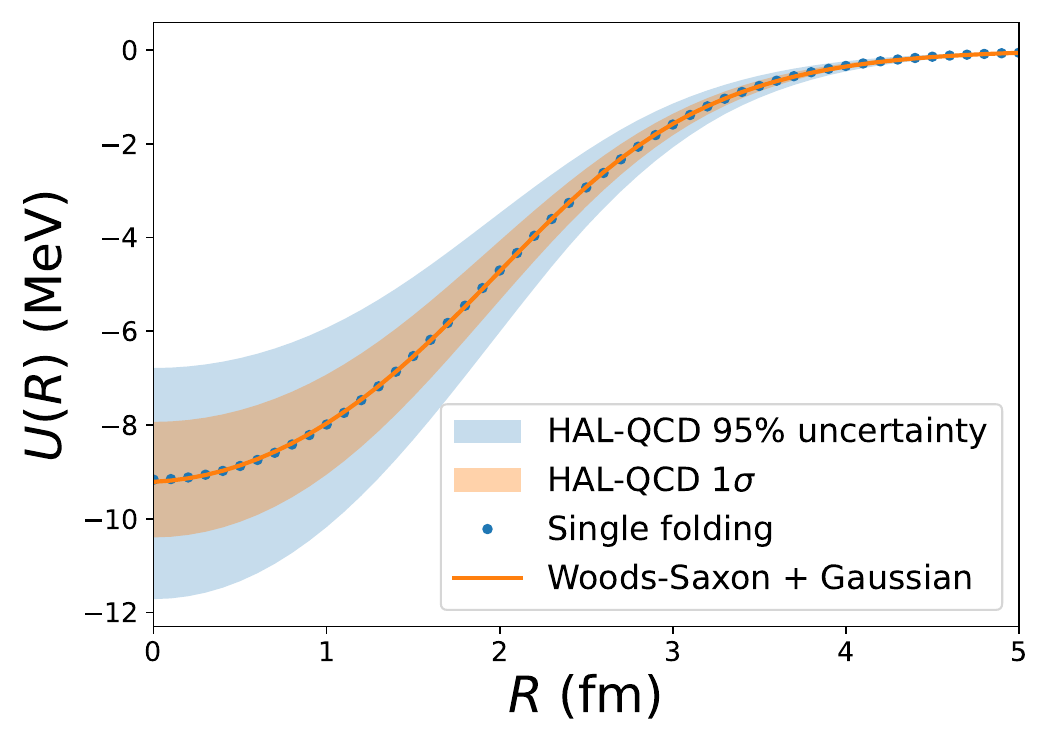}
	}
		\hfill
	\subfloat[$J/\psi N^{\textrm{spin-ave}}$]{
		\includegraphics[width=0.45\textwidth]
		{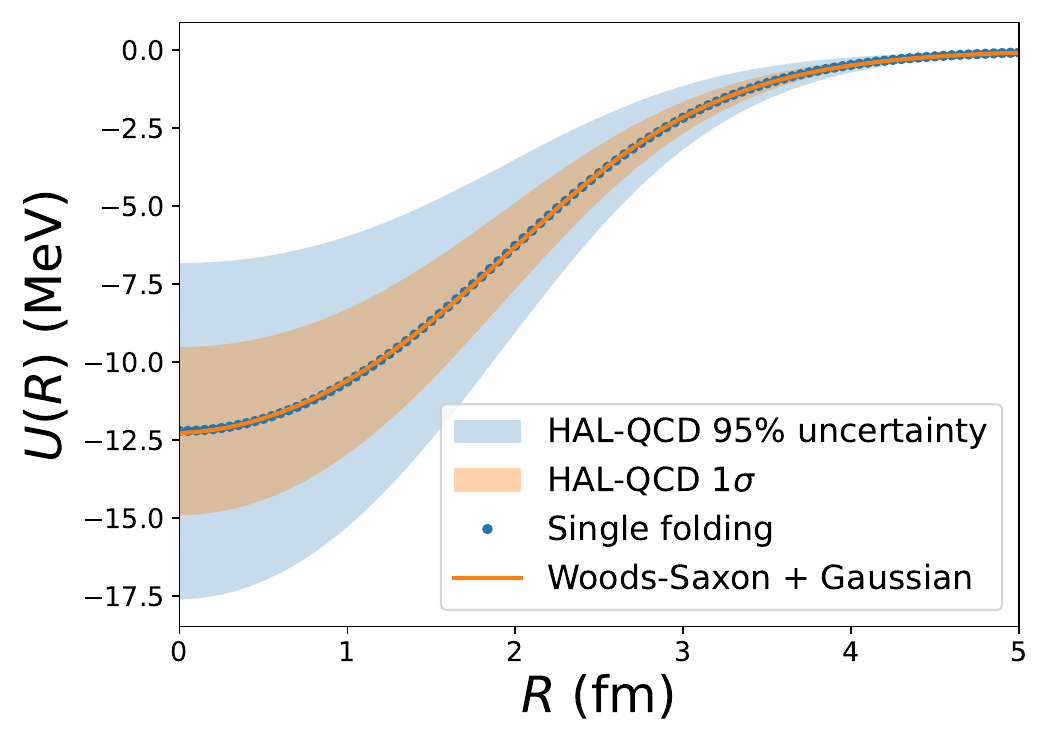}
	}
		\caption{
		Comparison between the numerical single-folding $^{10}$Be--$c\bar c$
		potentials and the Woods--Saxon plus Gaussian parametrizations for the
		three HAL QCD charmonium--nucleon channels. The point data show the
		folded potentials, where the symbols represent the numerical folding
		results, the solid curves denote the fitted Woods--Saxon plus Gaussian
		forms, and the shaded bands indicate the propagated statistical
		uncertainties from the HAL QCD interaction. 	}	
	\label{fig:folding_fit}	
\end{figure*}

The resulting parameterizations for the $J/\psi N$ ($^{4}S_{3/2}$), $J/\psi N$ ($^{2}S_{1/2}$), $\eta_cN$ ($^{2}S_{1/2}$) and $J/\psi N^{\textrm{spin-ave}}$ channels are compared with their respective numerical folding potentials in Fig.~\ref{fig:folding_fit}. In each panel, the symbols denote the numerical folding results, the solid curves represent the corresponding Woods--Saxon plus Gaussian parameterizations, while the shaded regions indicate the $68\%$ and $95\%$ confidence intervals obtained by propagating the statistical uncertainties of the HAL QCD interaction through the folding procedure.

The fitted Woods--Saxon plus Gaussian potentials accurately reproduce the numerical folding calculations over the entire radial interval. The maximum deviations are found to be $5.74\times10^{-2}$, $6.83\times10^{-2}$, $4.64\times10^{-2}$ and  $6.37\times10^{-2}$ MeV for the $J/\psi N$ ($^{4}S_{3/2}$), $J/\psi N$ ($^{2}S_{1/2}$), $\eta_cN$ ($^{2}S_{1/2}$) and $J/\psi N^{\textrm{spin-ave}}$ channels, respectively. The corresponding rms deviations are only $1.36\times10^{-2}$, $1.59\times10^{-2}$, $1.09\times10^{-2}$ and $1.54\times10^{-2}$ MeV, demonstrating that the analytical representation reproduces the numerical folding potentials with high fidelity.

The fitted parameters together with the corresponding fitting quality, volume integrals, and potential rms radii are summarized in Table~\ref{tab:folding-fit}. For all four channels, the fitting errors remain substantially smaller than the statistical uncertainties propagated from the underlying HAL QCD interaction. Therefore, the Woods--Saxon plus Gaussian parameterization introduces a negligible additional uncertainty while providing a computationally efficient representation of the folded interactions for the subsequent three-body calculations.

\begin{table*}[t]
	\centering
	\caption{Fitted parameters of the Woods--Saxon plus Gaussian representation of the folded
		$^{10}$Be--$c\bar c$ potentials for the four HAL QCD channels. The maximum fitting deviation
		($\Delta_{\rm max}$), rms fitting deviation ($\Delta_{\rm rms}$), volume integral ($J_V$),
		and potential rms radius ($R_{\rm rms}$) are also listed. The quoted uncertainties of the fitted
		parameters correspond to one standard deviation obtained from the covariance matrix of the
		nonlinear least-squares fit.}
	\label{tab:folding-fit}
	\begin{tabular}{lccccccc}
		\hline\hline
		Channel &
		$J/\psi N(^4S_{3/2})$ &
		$J/\psi N(^2S_{1/2})$ &
		$\eta_cN(^2S_{1/2}) $ &
		$J/\psi N^{\textrm{spin-ave}}$
		\\
		\hline
		
		$V_0$ (MeV)
		&
		5.363(718)
		&
		6.699(835)
		&
		4.636(573)
		& 
		5.820(608)
		\\
		
		$R_0$ (fm)
		&
		2.121(23)
		&
		2.112(22)
		&
		2.107(21)
		&  
		2.115(18)
		\\
		
		$a$ (fm)
		&
		0.542(24)
		&
		0.540(22)
		&
		0.538(22)
		&
		0.544(18)
		\\
		
		$V_G$ (MeV)
		&
		6.412(675)
		&
		7.496(785)
		&
		4.667(538)
		&
		6.572(571)
		\\
		
		$b_G$ (fm)
		&
		2.299(12)
		&
		2.306(11)
		&
		2.288(13)
		&
		2.300(10)
		\\
		
		\hline
		
		$\Delta_{\rm max}$ (MeV)
		&
		0.0574
		&
		0.0683
		&
		0.0464
		&
		0.0637
		\\
		
		$\Delta_{\rm rms}$ (MeV)
		&
		0.0136
		&
		0.0159
		&
		0.0109
		&
		0.0154
		\\
		
		$J_V$ (MeV\,fm$^3$)
		&
		-785.29
		&
		-945.88
		&
		-609.16
		&
		-825.59
		\\
		
		$R_{\rm rms}$ (fm)
		&
		2.725
		&
		2.723
		&
		2.699
		&
		2.722
		\\
		
		\hline\hline
	\end{tabular}
\end{table*}

It is noteworthy that the fitted geometric parameters ($R_0, a, b_G$) are nearly identical for all three channels, indicating that the spatial extension of the folded interaction is predominantly determined by the $^{10}$Be matter density. In contrast, the channel dependence is mainly reflected in the potential strengths ($V_{0}$ and $V_{G}$) which inherit the differences among the underlying HAL QCD charmonium--nucleon interactions.

To further characterize the fitted interactions beyond the potential parameters listed in Table~\ref{tab:folding-fit}, the corresponding two-body observables are calculated  by solving the radial Schr\"{o}dinger equation using both the numerical folding potentials and their Woods--Saxon plus Gaussian representations. The calculated observables include the $s$-wave binding energy ($B$) when a bound state exists,  the root-mean-square radius of the bound-state wave function ($r_{\rm rms}$) the scattering length ($a_0$) and the effective range ($r_e$) are presented in Table~\ref{tab:comparison}. These quantities provide physically transparent measures of the interaction strength and offer an independent validation of the analytical parameterization. The agreement obtained between the numerical folding potentials and the fitted analytical forms confirms that the latter accurately reproduce the low-energy properties relevant for the subsequent three-body calculations. The statistical uncertainties of the folding potentials, obtained from the HAL QCD analysis, are propagated to the calculated two-body observables by repeating the calculations using the central, upper, and lower uncertainty-band potentials.

The scattering length and effective range are calculated from the low-energy behavior of the $S$-wave phase shift, via the effective range expansion (ERE) formula up to the next-leading-order,		
\begin{equation}
	q\cot\delta_{0}=-\frac{1}{a_{0}}+\frac{1}{2}r_{e}q^{2}+\mathcal{O}\left(q^{4}\right).\label{eq:ERE}
\end{equation}	

For a generic two-body observable $X$, the uncertainty associated with the HAL QCD interaction is determined from the results obtained with the upper and lower uncertainty-band potentials. The resulting observable is expressed as
\begin{equation}
X=X_{0}{}_{-\left(X_{0}-X_{-}\right)}^{+\left(X_{+}-X_{0}\right)},
	\label{eq:uncertainty_propagation}
\end{equation}
where $X_0$ denotes the result obtained with the central potential, while $X_+$ and $X_-$ denote the corresponding results obtained with the upper and lower uncertainty-band potentials, respectively.

\begin{table*}[t]
	\centering
	\caption{
		Comparison of low-energy observables obtained from the numerical folding $^{10}$Be--$c\bar c$
		potentials and from their Woods--Saxon plus Gaussian representations.
	}
	\label{tab:comparison}
	\begin{tabular}{llcccc}
		\hline\hline
		
		Channel &Potential &$B$ (MeV) &$r_{\rm rms}$ (fm) &$a_0$ (fm) &$r_e$ (fm)
		\\
		\hline
		
		\multirow{2}{*}{$J/\psi N(^4S_{3/2})$}
		& Folding     & $2.319_{-1.365}^{+1.634}$ & $2.539_{-0.336}^{+0.762}$ & $2.163_{-0.262}^{+0.190}$ & $1.020_{-0.323}^{+0.205}$\\
		& WS+Gaussian & 2.318 & 2.538 & 2.164 & 1.036\\
		\hline
		
		\multirow{2}{*}{$J/\psi N(^2S_{1/2})$}
		& Folding     & $3.537_{-2.439}^{+2.962}$& $2.262_{-0.320}^{+0.882}$ & $2.297_{-0.391}^{+0.249}$ & $1.170_{-0.444}^{+0.242}$\\
        & WS+Gaussian & 3.535 & 2.263 & 2.297 & 1.186\\
		\hline
		
		\multirow{2}{*}{$\eta_cN(^2S_{1/2})$}
		& Folding     & $1.162_{-0.448}^{+0.510}$ & $3.128_{-0.324}^{+0.530}$ & $1.942_{-0.118}^{+0.102}$ & $0.753_{-0.164}^{+0.130}$\\ 
        & WS+Gaussian & 1.162 & 3.130 & 1.943 & 0.769\\	
        \hline
        
        \multirow{2}{*}{$J/\psi N^{\textrm{spin-ave}}$}
        & Folding     & $2.613_{-1.235}^{+1.412}$ & $2.455_{-0.265}^{+0.91}$ & $2.198_{-0.202}^{+0.156}$ & $1.066_{-0.239}^{+0.166}$\\ 
        & WS+Gaussian & 2.613 & 2.455 & 2.200 & 1.278\\		
		\hline\hline
	\end{tabular}
\end{table*}

\begin{figure*}[hbt!]
	\centering
	\includegraphics[scale=1.0]{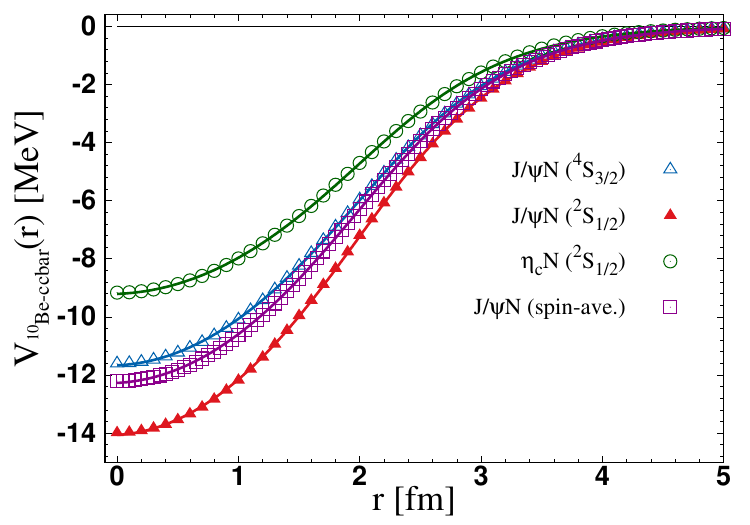}
			\caption{
				Comparison of the effective $^{10}\textrm{Be}\textrm{-}c\bar{c}$ single-folding potentials (data points) obtained from the $J/\psi N(^{4}S_{3/2})$ (blue triangles), $J/\psi N(^{2}S_{1/2})$ (filled red triangles), $\eta_cN(^{2}S_{1/2})$ (open green circles), and spin-averaged $J/\psi N$ (purple squares) interactions. The corresponding Woods--Saxon plus Gaussian fits are shown by solid lines. The fitted parameters are listed in Table~\ref{tab:comparison}.
		\label{fig:v_CharmAlpha} 
	}
\end{figure*}

Finally, to facilitate a direct comparison of the radial behavior of the four effective single-folding interactions, namely, $J/\psi\textrm{-}^{10}\textrm{Be}(^{4}S_{3/2})$, $J/\psi\textrm{-}^{10}\textrm{Be}(^{2}S_{1/2})$, $\eta_c\textrm{-}^{10}\textrm{Be}(^{2}S_{1/2})$, and the spin-averaged $J/\psi\textrm{-}^{10}\textrm{Be}$ interaction, a comparison of their radial behaviors is presented in Fig.~\ref{fig:v_CharmAlpha}. These effective $^{10}\textrm{Be}\textrm{-}c\bar{c}$ potentials are obtained by solving the single-folding equation, Eq.~\eqref{eq:folding}, using the corresponding $J/\psi N(^{4}S_{3/2})$, $J/\psi N(^{2}S_{1/2})$, $\eta_cN(^{2}S_{1/2})$, and spin-averaged $J/\psi N$ interactions as input. All four folded potentials are attractive over the relevant radial region, although their strengths and radial dependences exhibit noticeable channel dependence. In particular, the $J/\psi N(^{2}S_{1/2})$ interaction generates the strongest folded potential, whereas the $\eta_cN(^{2}S_{1/2})$ channel is comparatively weaker. The spin-averaged $J/\psi\textrm{-}^{10}\textrm{Be}$ potential lies between the two spin-dependent $J/\psi\textrm{-}^{10}\textrm{Be}$ interactions, as expected from its definition in terms of their spin-weighted combination. The corresponding Woods--Saxon plus Gaussian parametrizations provide an accurate representation of the numerical folding potentials and are used in the subsequent two- and three-body calculations.

In the two-body calculations, the experimental (physical) masses of the constituent particles were employed throughout. Specifically, $m_{J/\psi}=3096.900~\mathrm{MeV}/c^{2}$, $m_{\eta_c}=2983.9~\mathrm{MeV}/c^{2}$, and the nuclear mass $M(^{10}\mathrm{Be})=9324.945~\mathrm{MeV}/c^{2}$ were adopted. The corresponding reduced masses were found to be $\mu(J/\psi\textrm{-}^{10}\mathrm{Be})=2324.809~\mathrm{MeV}/c^{2}$ and $\mu(\eta_c\textrm{-}^{10}\mathrm{Be})=2260.545~\mathrm{MeV}/c^{2}$, respectively. Consequently, the kinetic-energy coefficients entering the radial Schr\"{o}dinger equation were obtained as $\hbar^{2}/(2\mu)=8.37445$ and $8.62288~\mathrm{MeV,fm^{2}}$, respectively.

The folding procedure has been successfully employed in Refs.~\cite{Filikhin_2005, HiyamaPRC2022, filikhin2024phihe, ETMINAN2025PLB139564, etminanPRCphiNalpha2025} to analyze \(\Lambda Nn\textrm{-}\alpha\), \(\Xi n\textrm{-}\alpha\alpha\), \(\phi\alpha\alpha\), and \(\phi n\textrm{-}\alpha\) systems, and in Ref.~\cite{filikhin2000alpha8be}, where the first \(0_{2}^{+}\) excited state of \({}^{12}\)C was described on the basis of the Faddeev equation. Agreement with experimental data and with calculations performed in different theoretical frameworks has been reported for the predictions of these works.

\subsection{Choice of $n\textrm{-}^{10}\textrm{Be}$ ($n$-core) interaction}
\label{sec:nBe_potential}
The neutron-$^{10}$Be interaction is an essential ingredient in the present
three-body calculation, since the weakly bound nature of the $^{11}$Be
subsystem strongly affects the spatial extension of the
$c\bar{c}+n+^{10}$Be wave function. 
In the present work, the single-channel, $l$-dependent Woods--Saxon potential developed by
Thompson \textit{et al.}~\cite{ThompsonPRC1996}, is adopted. This interaction was constructed to
reproduce the low-lying spectrum of the one-neutron halo nucleus
$^{11}\mathrm{Be}$. 
 This interaction was constructed to describe the low-lying spectrum of the one-neutron halo nucleus $^{11}$Be and, in particular, the well-known inversion of the $1s_{1/2}$ and $0p_{1/2}$ levels.

The neutron--core potential is written as $V_{n-^{10}\mathrm{Be}}\left(r\right)=\ensuremath{V_{c}\left(r\right)}+V_{so}\left(r\right)\boldsymbol{l}\cdot\boldsymbol{s}$,
includes central and spin-orbit terms, both expressed as Woods–Saxon potentials,
\begin{equation}
V_{c}\left(r\right)=\frac{v_{c}^{\left(l\right)}}{1+\exp\left(\frac{r-R_{c}}{a_{c}}\right)},
	\label{eq:nBe_central}
\end{equation}
and the spin--orbit contribution is given by
\begin{equation}
V_{so}\left(r\right)=-\left(\frac{\hbar}{m_{\pi}c}\right)^{2}\:\frac{2v_{so}^{\left(l\right)}}{ra_{c}^{\left(l\right)}}\frac{\exp\left(\frac{r-R_{c}}{a_{c}}\right)}{\left[1+\exp\left(\frac{r-R_{c}}{a_{c}}\right)\right]^{2}}.
	\label{eq:nBe_ls}
\end{equation}

Deep neutron-core potentials such as given in~\cite{NUNES199643,NUNES2002593,ThompsonPRC1996} can generate additional Pauli-forbidden bound
states corresponding to configurations already occupied in the $^{10}$Be
core. In conventional few-body calculations, such states are usually treated
by projection techniques or by constructing phase-equivalent potentials.
Here, a different approach is followed and modify only the Woods--Saxon depths,
keeping the radial geometry unchanged. This minimal modification eliminates
the spurious deeply bound states while preserving the physical properties of
the neutron-core system.

The geometry parameters of the Thompson interaction are retained in the
present calculation as
$	R_c=r_c A^{1/3}$,
with
$r_c=1.27~{\rm fm}$, and $ a_c=0.67~{\rm fm}$, which gives $R_c=1.27\times10^{1/3}=2.735~{\rm fm}$.

However, the depths of the Woods--Saxon terms are
 modified in order to remove the unphysical deeply bound
$s$-wave solutions and to reproduce the experimentally known low-energy
spectrum of $^{11}$Be. 
The negative-parity component of the neutron--core interaction is adjusted to reproduce the experimentally observed first excited $0p_{1/2}$ state of $^{11}$Be, with a binding energy of $E(0p_{1/2})=-0.183~\mathrm{MeV}$. For the positive-parity component, the central potential strength is modified to reproduce the weakly bound $1s_{1/2}$ ground state, with $E(1s_{1/2})=-0.504~\mathrm{MeV}$. Both energies are measured relative to the $n+{}^{10}$Be breakup threshold.

The expectation value of the spin--orbit operator, $\langle\boldsymbol{l}\cdot\boldsymbol{s}\rangle$,
for the $0p_{1/2}$ and $1s_{1/2}$ states, gives
$-1$ and $0$,
respectively. The numerical value of the factor appearing in the spin--orbit term is 
$\left({\hbar}/{m_{\pi}c}\right)^{2} = 1.9989$ $\textrm{fm}^{2}$ $\thickapprox2.00 $ $\textrm{fm}^{2}$. 
The parameters of the original Thompson neutron-$^{10}$Be interaction and
the modified interaction used in the present work are given in Table~\ref{tab:nBe_parameters}.

\begin{table}[ht]
	\centering
	\caption{
		Comparison between the original Thompson neutron-$^{10}$Be interaction and
		the modified interaction used in the present work. The Woods--Saxon geometry
		parameters are kept unchanged, while only the potential depths  (in MeV) are adjusted
		to reproduce the low-lying spectrum of $^{11}$Be.
	}
	\label{tab:nBe_parameters}
	\begin{tabular}{c c c }
		\hline\hline
		 Parameter & Thompson~\cite{ThompsonPRC1996} & Present modified potential \\
		\hline		
		$v_{c}^{\left(s\right)}$ 
		& $-55.06$
		& $-10.10$ \\
		
		$v_{c}^{\left(p\right)}$ 
		& $-36.250$
		& $-24.24$ \\
		
		$v_{c}^{\left(d\right)}$ 
		& $-60.03$
		& $-10.10$ \\
		
		$v_{so}^{\left(p\right)}$ 
		& $+6.43$
		& $+6.43$ \\
		
		$v_{so}^{\left(d\right)}$ 
		& $0.0$
		& $-6.43$ \\
		\hline
		
		\multicolumn{3}{c}{
			$R_{c}=2.736$ fm,\qquad $a_{c}=0.67$ fm
		}\\
		
		\hline\hline
	\end{tabular}
\end{table}

To verify the reliability of the adopted neutron-core potential, the
two-body Schr\"odinger equation for the $n$-$^{10}$Be system was solved
independently. The binding energy, neutron-core rms separation, matter radius,
charge radius, and asymptotic normalization coefficient (ANC) were calculated
and compared with available experimental information or or extracted from reaction analyses. The neutron-core rms
radius is obtained from the normalized radial wave function as
\begin{equation}
	r_{n-^{10}{\rm Be}}
	=
	\left[
	\int_0^\infty r^2 u^2(r)dr
	\right]^{1/2},
\end{equation}
where $u(r)$ is the reduced radial wave function.
The matter radius of $^{11}$Be was evaluated using the relation
\begin{equation}
	R_m^2(^{11}{\rm Be})
	=
	\frac{10}{11}R_m^2(^{10}{\rm Be})
	+
	\frac{10}{121}r_{n-^{10}{\rm Be}}^2 ,
\end{equation}
where the matter radius of the $^{10}$Be core was taken as $R_m(^{10}{\rm Be})=2.30$ fm~\cite{TANIHATA1988592}. 
The calculated neutron separation energy is
\begin{equation}
	S_n(^{11}{\rm Be})=0.505~{\rm MeV},
\end{equation}
which is in good agreement with the experimental value
$S_n=0.504(6)$ MeV~\cite{AUDI199366}. The calculated neutron-core rms radius,
\begin{equation}
	r_{n-^{10}{\rm Be}}=6.145~{\rm fm},
\end{equation}
together with the small asymptotic wave number,
\begin{equation}
	\kappa=
	\sqrt{\frac{|E|}{\hbar^2/(2\mu)}}
	=0.151~{\rm fm}^{-1},
\end{equation}
demonstrates the extended halo structure of the $^{11}$Be ground state.

The asymptotic normalization coefficient was extracted by fitting the
large-distance behavior of the radial wave function according to
$	u(r)\rightarrow C\exp(-\kappa r),$
which gives
$	C=0.74~{\rm fm}^{-1/2}.$
The calculated observables are summarized in Table~\ref{tab:nBe_validation}.

\begin{table}[ht]
	\centering
	\caption{
		Comparison between calculated properties of the $^{11}$Be ground state using
		the modified Thompson neutron-$^{10}$Be interaction and experimental values (or extracted from reaction analyses).
	}
	\label{tab:nBe_validation}
	\begin{tabular}{c c c}
		\hline\hline
		Observable & Calculation$\:\:$ &  Exp.(or Reference value) \\
		\hline
		$S_n(^{11}{\rm Be})$ (MeV)
		& 0.505 & 0.504 \\
		
		$r_{n-^{10}{\rm Be}}$ (fm)
		& 6.145 & 5.7(4)~\cite{PhysRevC.68.034318} \\
		
		$R_m(^{11}{\rm Be})$ (fm)
		& 2.816 & 2.734(5)~\cite{TANIHATA1988592}, 2.90(5)~\cite{PhysRevLett.76.3903} \\
		
		$R_{\rm ch}(^{11}{\rm Be})$ (fm)
		& 2.422 & 2.463(16)~\cite{PhysRevLett.102.062503}\\
		
		$\kappa$ (fm$^{-1}$)
		& 0.151 & 0.1483~\cite{AUDI199366,Wang_2021} \\
		
		ANC (fm$^{-1/2}$)
		& 0.74 & 0.75(4)~\cite{PhysRevC.78.011601}, 0.71(5)~\cite{PhysRevLett.108.192701}\\
		
		\hline\hline
	\end{tabular}
\end{table}

The good agreement between the calculated neutron separation energy and the available halo observables supports the use of the modified neutron--$^{10}$Be interaction as a realistic description of the $^{11}$Be subsystem. The interaction is therefore employed as the neutron--core component of the subsequent three-body calculations.

For the three-body calculations, the neutron--$^{10}$Be interaction is retained in the partial waves relevant to the low-energy structure of the $^{11}$Be subsystem. In particular, the $s$-, $p$-, and $d$-wave components are included, corresponding to the $1s_{1/2}$ bound state, the $0p_{1/2}$ excited state, and the low-lying $d$-wave configurations of $^{11}$Be, respectively. The $l$ dependence of the neutron--core interaction is therefore explicitly retained in the three-body Hamiltonian. Higher partial waves were examined and found to make negligible contributions to the low-energy states considered here.

The same neutron--$^{10}$Be interaction is used consistently in all subsequent three-body calculations.
 \section{Three-cluster systems} \label{Three-cluster systems}		
The numerical analysis of the ground-state central binding energy, denoted by \(B_3\), and the nuclear matter radius, \(r_{\text{mat}}\), for the three-body system \(c\bar{c}+n+^{10}\textrm{Be}\) is presented and discussed in this section. The coupled equations given in Eqs.~\eqref{eq:Faddeev_{c}oupled} and the angular integrations in Eq.~\eqref{eq:V_ij} are solved in accordance with the procedures implemented in the FaCE code~\cite{face}. The two-body interaction parameters utilized in these computations correspond to those described in Sec.~\ref{sec:Two-body-potentials}.  

\subsection{Convergence with respect to $K_{\max}$}
\begin{figure*}[t]
	\centering
	\includegraphics[width=0.48\textwidth]{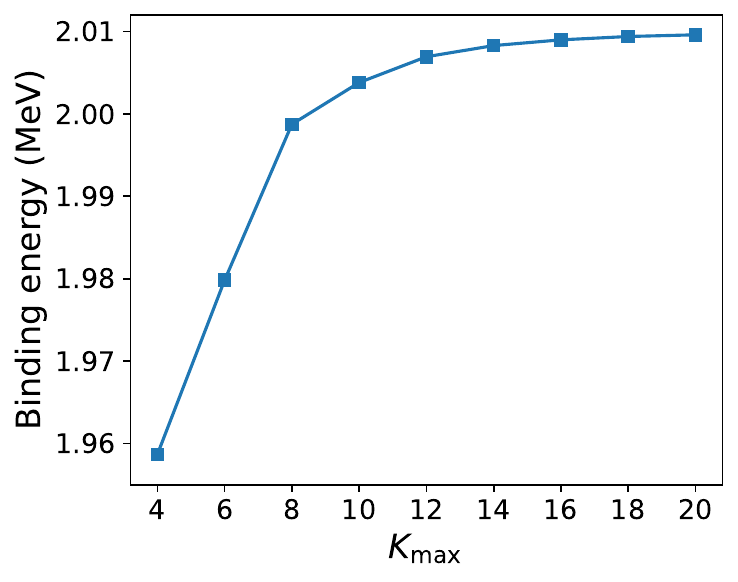}
	\hfill
	\includegraphics[width=0.48\textwidth]{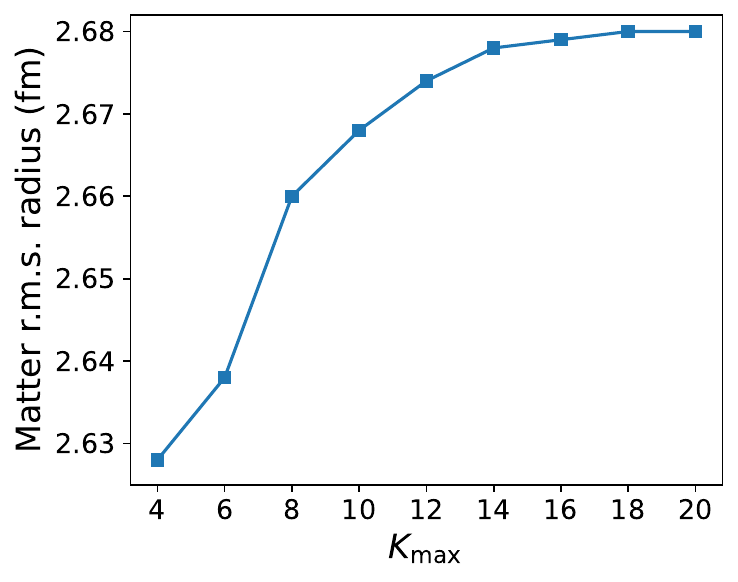}
	\caption{
		Convergence of the hyperspherical harmonics calculations for the
		$\eta_c+n+{}^{10}\mathrm{Be}$ system with $ j = 1/2 $ as a function of the
		maximum hypermomentum $K_{\max}$.
		Left panel: three-body binding energy.
		Right panel: matter root-mean-square (r.m.s.) radius.
		Both observables exhibit excellent convergence with increasing
		$K_{\max}$, and the adopted value $K_{\max}=20$ provides numerically
		converged results. Calculations use $i_{\textrm{max}} = 20$.
	}
	\label{fig:kconv}
\end{figure*}

The numerical convergence of the hyperspherical harmonics expansion was
examined for the $\eta_c+n+{}^{10}\mathrm{Be}$ system with total spin
$j=1/2$, which represents the weakest bound state considered in the
present work and therefore provides the most stringent convergence test.
Figure~\ref{fig:kconv} shows the dependence of the three-body binding
energy and the matter r.m.s. radius on the maximum hypermomentum
$K_{\max}$.

As $K_{\max}$ increases, both observables exhibit rapid convergence. The
variation of the binding energy decreases from several tens of keV at
small $K_{\max}$ to less than $1$~keV for $K_{\max}\ge16$. In
particular, the difference between the $K_{\max}=18$ and $20$
calculations is only $0.20$~keV, while the matter r.m.s. radius remains
unchanged within $0.001$~fm. These results demonstrate that the adopted
value $K_{\max}=20$ provides numerically converged three-body
observables.

\subsection{Three-body binding energies, uncertainty propagation, and core excitation}
\label{sec:three-body-results}

Table~\ref{tab:Charm-alpha-para} summarizes the calculated two- and three-body observables for the three spin-dependent charmonium--nucleon interactions. The central two-body binding energies of the effective $^{10}$Be--$c\bar{c}$ interactions range from $1.162$ to $3.537$ MeV, while the corresponding three-body calculations yield bound states with binding energies between $2.009$ and $4.650$ MeV. The uncertainties quoted for the three-body binding energies and matter radii are obtained by propagating the statistical uncertainties of the fitted Woods--Saxon plus Gaussian parameters of the effective $^{10}$Be--$c\bar{c}$ interactions, as described in Sec.~\ref{sec:folding-Model}.

The propagated uncertainties remain moderate for the calculated three-body binding energies, whereas the matter radii are comparatively less sensitive to variations in the effective interaction. This difference reflects the stronger dependence of the binding energy on the strength and detailed shape of the interaction, while the spatial extension of the three-body system is comparatively more stable. For the three spin-dependent interactions, the calculated matter radii lie in the range $2.474$--$2.680$ fm, with the largest radius corresponding to the least strongly bound $\eta_c+n+{}^{10}$Be system.

For comparison, the lower part of Table~\ref{tab:Charm-alpha-para} presents the corresponding results obtained using the spin-averaged $J/\psi$--$^{10}$Be interaction. This interaction is constructed by folding the spin-averaged $J/\psi N$ potential defined in Eq.~\eqref{eq:NJpsi-spin-ave}. The resulting binding energies for $j=3/2$ and $j=1/2$ are $3.606(428)$ and $3.688(430)$ MeV, respectively. These values are close to each other, indicating that the spin-averaged interaction largely removes the dependence on the total angular momentum associated with the spin dependence of the underlying $J/\psi N$ interaction. The corresponding matter radii, $2.528(18)$ and $2.506(18)$ fm, are also very similar.

In the present calculations, the $^{10}$Be core is treated in a two-state coupled-channel approximation that explicitly includes its $0_1^+$ ground state and first excited $2_1^+$ state. The excitation energy of the latter is taken as $E_x(2_1^+)=3.368$ MeV, and the coupling between the core states is described using the deformed-core prescription with a deformation length of $1.6638$ fm. Consequently, the three-body wave function contains contributions from both core configurations. The three Faddeev components corresponding to the three Jacobi coordinate sets are retained, and the allowed orbital angular momenta are specified by $l_x,l_y,l_t\leq2$, together with the corresponding spin and total-angular-momentum couplings.

Although the excited-core component is explicitly included, its contribution to the calculated ground states is small. The probability of finding the $^{10}$Be core in its $2_1^+$ state is approximately $1.2\%$, $1.55\%$, and $0.79\%$ for the $J/\psi N(^4S_{3/2})$, $J/\psi N(^2S_{1/2})$, and $\eta_cN(^2S_{1/2})$ interactions, respectively. The corresponding probabilities for the spin-averaged $J/\psi N$ interaction are also small, approximately $1.27\%$ and $1.30\%$ for $j=3/2$ and $j=1/2$, respectively. Thus, the calculated ground states are predominantly associated with the $^{10}$Be ground-state configuration, while the explicit inclusion of the $2_1^+$ excitation provides a consistent treatment of core polarization within the adopted coupled-channel model.

\begin{table}[hbt!]
	\centering
		\caption{
			Central two-body binding energies of the $^{10}$Be--$c\bar{c}$ systems, $B_{^{10}\mathrm{Be}-c\bar{c}}$, together with the calculated ground-state three-body binding energies, $B_3$, nuclear matter radii, $r_{\mathrm{mat}}$, and probabilities of the $^{10}$Be $2_1^+$ core configuration, $P(^{10}\mathrm{Be},2_1^+)$, for the corresponding $c\bar{c}+n+{}^{10}$Be systems. The values given in parentheses denote the propagated statistical uncertainties arising from the uncertainties of the fitted Woods--Saxon plus Gaussian parameters of the effective $^{10}$Be--$c\bar{c}$ interaction, as described in Sec.~\ref{sec:folding-Model}. The lower part of the table presents the corresponding results obtained using the spin-averaged $J/\psi$--$^{10}$Be interaction.
			\label{tab:Charm-alpha-para}}
			\begin{tabular}{cccccccc}
		\hline
		\hline 
	 System & $ j $ &$B_{^{10}\textrm{Be}\textrm{-}c\bar{c}}$ (MeV) && $B_3$ (MeV) & $r_{\text{mat}}$ (fm)&&$P(^{10}\mathrm{Be},2_1^+)$ \\
		\hline
 \multirow{2}{*}{ $J/\psi+n+^{10}$Be} 
   & 3/2 & 2.319 && 3.297(493) & 2.542(24)&& $\sim1.2 \%$ \\
   & 1/2 & 3.537 && 4.650(632) & 2.474(17)&&$ \sim1.55\%$ \\
 $\eta_{c}+n+^{10}$Be
   & 1/2 & 1.162 && 2.009(325) & 2.680(42)&& $\sim0.79\% $\\
 \hline
 	\multicolumn{8}{c}{ 	Using spin-ave $J/\psi\textrm{-}^{10}$Be potential}\\
 	 \cline{2-6}
  \multirow{2}{*}{ $J/\psi+n+^{10}$Be}
   &  3/2 & 2.613 && 3.606(428) & 2.528(18)&&$\sim1.27\%$\\ 
   &  1/2 & 2.613 && 3.688(430) & 2.506(18)&&$\sim1.30\%$\\ 
		\hline
		\hline 	
	\end{tabular}
\end{table}

To estimate the statistical uncertainties associated with the effective $c\bar{c}$--$^{10}$Be interaction, the statistical uncertainties of the fitted Woods--Saxon plus Gaussian potential parameters are propagated to the calculated three-body observables. The effective interaction is characterized by the five parameters $(V_0,R_0,a,V_G,b_G)$, for which the best-fit values and one-standard-deviation uncertainties are obtained from fits to the single-folding potential.

The uncertainty propagation is performed using a finite-difference sensitivity analysis. Starting from the central parameter set, each parameter is varied independently by its corresponding $1\sigma$ uncertainty while all remaining parameters are kept fixed. For every parameter $p_i$, two additional three-body calculations are carried out using $p_i+\sigma_i$ and $p_i-\sigma_i$, respectively. Consequently, for a potential described by five fitted parameters, a total of eleven three-body calculations are performed, consisting of one calculation for the central parameter set and ten calculations corresponding to the positive and negative variations of each parameter.

The uncertainty contribution of the parameter $p_i$ to an observable $O$ (binding energy or matter radius) is estimated as
\begin{equation}
	\Delta O_i=\frac{|O_i^{(+)}-O_i^{(-)}|}{2},
\end{equation}
where $O_i^{(+)}$ and $O_i^{(-)}$ denote the values obtained with the positive and negative variations of the parameter $p_i$, respectively, while all remaining parameters are kept fixed. Following the HAL QCD analysis, the fitted parameters are treated as statistically independent~\cite{LyuPLB2025}. The total propagated uncertainty is then evaluated by adding the individual contributions in quadrature,
\begin{equation}
	\sigma_O=
	\left[
	\sum_i
	(\Delta O_i)^2
	\right]^{1/2},
\end{equation}
where the summation extends over the five fitted potential parameters.

This procedure quantifies the sensitivity of the three-body predictions to the statistical uncertainties of the fitted effective interaction and provides an estimate of the propagated uncertainties of the calculated binding energies and matter radii. The present analysis accounts only for the statistical uncertainties of the fitted interaction parameters and does not include additional systematic uncertainties associated with the underlying HAL QCD interaction, the single-folding procedure, or the three-body model.

Table~\ref{tab:total_norm} summarizes the dominant hyperspherical-harmonics (HH) channels contributing to the ground-state wave functions of the $c\bar{c}+n+{}^{10}$Be systems. The quantity $P^{\beta K}$ denotes the probability (weight) of a given HH channel in the normalized three-body wave function, where each channel is specified by the quantum numbers $j, K, l_x, l_y, l, S_x$. Only channels with contributions exceeding $1\%$ are listed. The total angular momentum $j$ refers to the $c\bar{c}+n+{}^{10}$Be system in the Jacobi-$T$ coordinate set shown in Fig.~\ref{fig:T_jacobi}.

For all systems considered, the ground-state wave function is overwhelmingly dominated by the lowest hyperspherical configuration with $K=0$ and $(l_x,l_y)=(0,0)$, whose probability exceeds $93\%$ in every case. This indicates that the ground states are predominantly described by a configuration with zero orbital angular momentum in both Jacobi coordinates. The remaining probability is distributed among higher-$K$ channels, each contributing only at the percent level or less, with small admixtures involving nonzero orbital angular momenta. Thus, although higher HH components are fully retained in the calculation, they provide relatively small corrections to the dominant configuration.

The pronounced dominance of the lowest HH channel is also consistent with the convergence of the hyperspherical expansion with increasing $K_{\max}$. Consequently, the calculated binding energies and matter radii are largely determined by the leading $K=0$ configuration, while the higher-$K$ components provide smaller corrections. The complete converged HH expansion is nevertheless retained in the final calculations.

\begin{table}[hbt!]
	\centering
	\caption{
		Dominant HH channels contributing to the ground-state wave functions of the $c\bar{c}+n+{}^{10}$Be systems. The quantity $P^{\beta K}$ denotes the probability (weight) of each HH channel in the normalized wave function. Only channels with probabilities greater than $1\%$ are listed; the remaining probability is distributed among channels with individual contributions below $1\%$.
	}
	\label{tab:total_norm}
	\begin{tabular}{ccccccccc}
		\hline\hline
         System &$j$&$K$&$l_{x}$&$l_{y}$ &$l$ &$S_{x}$ && $P^{\beta K} $ 	\\
		\hline
		
		\multirow{2}{*}{ $J/\psi+n+^{10}$Be}
		
		&\multirow{2}{*}{$3/2$}& 0 & 0 & 0 & 0 & 1.5 && 95.0\\
		& & 4 & 1 & 1 & 0 & 1.5 &&  1.1\\
		\hline
		
		\multirow{4}{*}{ $J/\psi+n+^{10}$Be}
		&\multirow{4}{*}{$1/2$}& 0 & 0 & 0 & 0 & 0.5 && 93.8 \\
        && 2 & 1 & 1 & 0 & 0.5 && 1.1 \\
        && 2 & 0 & 0 & 0 & 0.5 && 1.3 \\
        && 4 & 1 & 1 & 0 & 0.5 && 1.1 \\
		\hline 
		
		\multirow{3}{*}{ $\eta_{c}+n+^{10}$Be}
		&\multirow{3}{*}{$1/2$}& 0 & 0 & 0 & 0 & 0.5 && 95.7\\
        && 4 & 2 & 2 & 0 & 0.5 && 1.1\\	
        && 4 & 1 & 1 & 0 & 0.5 && 1.2\\
		\hline\hline
	\end{tabular}
\end{table}

The creation and detection of such systems present significant experimental challenges. However, potential observations of these \(c\bar{c}+n+^{10}\textrm{Be}\) states could be achievable through hadron-beam experiments at facilities including Jefferson Lab, J-PARC, and FAIR, or via relativistic heavy-ion collisions at RHIC and the LHC.

\section{Summary and outlook\label{sec:Summary-and-conclusions}}
In this work, the possibility of forming a bound charmonium--nuclear system in the three-body configuration $c\bar c+n+{}^{10}$Be has been investigated. The system has been treated within a three-cluster model consisting of a charmonium cluster, a neutron, and a $^{10}$Be core described in a two-state coupled-channel approximation that includes its $0_1^+$ ground state and first excited $2_1^+$ state. The coupled Faddeev equations have been solved using a hyperspherical-harmonics expansion. The study has been aimed at connecting the charmonium--nucleon interactions determined from lattice QCD with a realistic few-body description of a bound neutron-rich nuclear subsystem.

Particular attention was devoted to the two-body interactions entering the three-body calculation. The $n$--$^{10}$Be interaction was constructed from the $l$-dependent Woods--Saxon interaction of Thompson \textit{et al.}~\cite{ThompsonPRC1996}. The Woods--Saxon geometry was retained, while the potential strengths were modified to eliminate deeply bound Pauli-forbidden states and to reproduce the low-energy properties of $^{11}$Be. An independent two-body calculation demonstrated that the resulting interaction gives a neutron separation energy, neutron--core rms separation, matter and charge radii, asymptotic wave number, and ANC consistent with the available experimental and reaction-analysis information. This provides a physically constrained description of the weakly bound $n+{}^{10}$Be subsystem used in the three-body calculation.

The effective $^{10}$Be--$c\bar c$ interactions were constructed by folding the $S$-wave $J/\psi N(^4S_{3/2})$, $J/\psi N(^2S_{1/2})$, and $\eta_cN(^2S_{1/2})$ interactions obtained by the HAL QCD Collaboration at nearly physical pion mass with the $^{10}$Be matter density. A spin-averaged $J/\psi N$ interaction was also considered. The numerical folding potentials were accurately represented by Woods--Saxon plus Gaussian parametrizations, with rms fitting deviations of only $0.011$--$0.016$ MeV. The resulting two-body observables, including binding energies, scattering lengths, and effective ranges, were found to be consistently reproduced by the analytical parametrizations. Statistical uncertainties of the HAL QCD input were propagated through the folding and fitting procedures and subsequently to the three-body observables.

All four effective interactions considered in the present study generate a bound $c\bar c+n+{}^{10}$Be system. For the spin-dependent interactions, the calculated central three-body binding energies are
$ B_3=3.297(493)\ {\rm MeV} $ for $J/\psi N(^4S_{3/2})$,
$ B_3=4.650(632)\ {\rm MeV} $ for $J/\psi N(^2S_{1/2})$, and
$ B_3=2.009(325)\ {\rm MeV} $ for $\eta_cN(^2S_{1/2})$. 
The corresponding nuclear matter rms radii are $2.542(24)$, $2.474(17)$, and $2.680(42)$ fm, respectively. For the spin-averaged $J/\psi N$ interaction, binding energies of $3.606(428)$ and $3.688(430)$ MeV are obtained for total angular momenta $j=3/2$ and $j=1/2$, respectively, with matter radii of $2.528(18)$ and $2.506(18)$ fm. These results demonstrate a clear sensitivity of the binding energy to the short-distance spin dependence of the underlying charmonium--nucleon interaction, while the spatial extension of the three-body system varies more moderately.

The numerical reliability of the three-body calculation was tested explicitly through the convergence of the hyperspherical expansion. Using the weakest-bound $\eta_c+n+{}^{10}$Be system as the most stringent test, both the binding energy and matter rms radius were found to be stable with increasing $K_{\max}$, with changes below $1$ keV in the binding energy for $K_{\max}\geq16$. The adopted value $K_{\max}=20$ therefore provides numerically converged three-body observables.  The dominant $K=0$ hyperspherical configuration accounts for $93.8\%$--$95.7\%$ of the ground-state wave function, depending on the interaction. Thus, the calculated states are strongly dominated by configurations with relative $s$-wave motion, while higher-$K$ components provide only small corrections. The rapid convergence and the dominance of the lowest hyperspherical component provide additional support for the numerical stability and the simple spatial structure of the predicted states.

Overall, the present results show that the attractive charmonium--nucleon interactions obtained from lattice QCD are sufficiently strong to generate bound $c\bar c+n+{}^{10}$Be configurations within the present three-body model. The predicted binding energies, however, carry sizeable statistical uncertainties associated with the effective $^{10}$Be--$c\bar c$ interaction and exhibit a significant dependence on the spin channel. The quoted uncertainties include the statistical uncertainties of the fitted effective interaction but do not include possible systematic effects associated with the HAL QCD extraction, the single-folding approximation, or the inert-core three-body model.

Further progress will require improved constraints on the charmonium--nucleon interaction and a systematic investigation of these additional sources of uncertainty. Extensions beyond the inert-core approximation, including explicit excited-core configurations and additional many-body dynamics, would provide a more complete description of the $c\bar c+n+{}^{10}$Be system. Experimental information on charmonium--nuclear systems would ultimately provide an important test of the QCD-based interactions and of the few-body description developed here.
		
\bibliography{Refs.bib}

\begin{thebibliography}{50}%
\makeatletter
\providecommand \@ifxundefined [1]{%
 \@ifx{#1\undefined}
}%
\providecommand \@ifnum [1]{%
 \ifnum #1\expandafter \@firstoftwo
 \else \expandafter \@secondoftwo
 \fi
}%
\providecommand \@ifx [1]{%
 \ifx #1\expandafter \@firstoftwo
 \else \expandafter \@secondoftwo
 \fi
}%
\providecommand \natexlab [1]{#1}%
\providecommand \enquote  [1]{``#1''}%
\providecommand \bibnamefont  [1]{#1}%
\providecommand \bibfnamefont [1]{#1}%
\providecommand \citenamefont [1]{#1}%
\providecommand \href@noop [0]{\@secondoftwo}%
\providecommand \href [0]{\begingroup \@sanitize@url \@href}%
\providecommand \@href[1]{\@@startlink{#1}\@@href}%
\providecommand \@@href[1]{\endgroup#1\@@endlink}%
\providecommand \@sanitize@url [0]{\catcode `\\12\catcode `\$12\catcode
  `\&12\catcode `\#12\catcode `\^12\catcode `\_12\catcode `\%12\relax}%
\providecommand \@@startlink[1]{}%
\providecommand \@@endlink[0]{}%
\providecommand \url  [0]{\begingroup\@sanitize@url \@url }%
\providecommand \@url [1]{\endgroup\@href {#1}{\urlprefix }}%
\providecommand \urlprefix  [0]{URL }%
\providecommand \Eprint [0]{\href }%
\providecommand \doibase [0]{https://doi.org/}%
\providecommand \selectlanguage [0]{\@gobble}%
\providecommand \bibinfo  [0]{\@secondoftwo}%
\providecommand \bibfield  [0]{\@secondoftwo}%
\providecommand \translation [1]{[#1]}%
\providecommand \BibitemOpen [0]{}%
\providecommand \bibitemStop [0]{}%
\providecommand \bibitemNoStop [0]{.\EOS\space}%
\providecommand \EOS [0]{\spacefactor3000\relax}%
\providecommand \BibitemShut  [1]{\csname bibitem#1\endcsname}%
\let\auto@bib@innerbib\@empty
\bibitem [{\citenamefont {Brodsky}\ \emph {et~al.}(1990)\citenamefont
  {Brodsky}, \citenamefont {Schmidt},\ and\ \citenamefont
  {de~T\'eramond}}]{BrodskyPRL1990}%
  \BibitemOpen
  \bibfield  {author} {\bibinfo {author} {\bibfnamefont {S.~J.}\ \bibnamefont
  {Brodsky}}, \bibinfo {author} {\bibfnamefont {I.}~\bibnamefont {Schmidt}},\
  and\ \bibinfo {author} {\bibfnamefont {G.~F.}\ \bibnamefont
  {de~T\'eramond}},\ }\bibfield  {title} {\bibinfo {title} {{Nuclear-bound
  quarkonium}},\ }\href {https://doi.org/10.1103/PhysRevLett.64.1011}
  {\bibfield  {journal} {\bibinfo  {journal} {Phys. Rev. Lett.}\ }\textbf
  {\bibinfo {volume} {64}},\ \bibinfo {pages} {1011} (\bibinfo {year}
  {1990})}\BibitemShut {NoStop}%
\bibitem [{\citenamefont {Voloshin}(2008)}]{VOLOSHIN2008455}%
  \BibitemOpen
  \bibfield  {author} {\bibinfo {author} {\bibfnamefont {M.}~\bibnamefont
  {Voloshin}},\ }\bibfield  {title} {\bibinfo {title} {Charmonium},\ }\href
  {https://doi.org/https://doi.org/10.1016/j.ppnp.2008.02.001} {\bibfield
  {journal} {\bibinfo  {journal} {Prog. Part. Nucl. Phys.}\ }\textbf {\bibinfo
  {volume} {61}},\ \bibinfo {pages} {455} (\bibinfo {year} {2008})}\BibitemShut
  {NoStop}%
\bibitem [{\citenamefont {Tsushima}\ \emph {et~al.}(2011)\citenamefont
  {Tsushima}, \citenamefont {Lu}, \citenamefont {Krein},\ and\ \citenamefont
  {Thomas}}]{KreinPRC2011}%
  \BibitemOpen
  \bibfield  {author} {\bibinfo {author} {\bibfnamefont {K.}~\bibnamefont
  {Tsushima}}, \bibinfo {author} {\bibfnamefont {D.~H.}\ \bibnamefont {Lu}},
  \bibinfo {author} {\bibfnamefont {G.}~\bibnamefont {Krein}},\ and\ \bibinfo
  {author} {\bibfnamefont {A.~W.}\ \bibnamefont {Thomas}},\ }\bibfield  {title}
  {\bibinfo {title} {{$J/\ensuremath{\Psi}$-nuclear bound states}},\ }\href
  {https://doi.org/10.1103/PhysRevC.83.065208} {\bibfield  {journal} {\bibinfo
  {journal} {Phys. Rev. C}\ }\textbf {\bibinfo {volume} {83}},\ \bibinfo
  {pages} {065208} (\bibinfo {year} {2011})}\BibitemShut {NoStop}%
\bibitem [{\citenamefont {Wasson}(1991)}]{WassonPRL1991}%
  \BibitemOpen
  \bibfield  {author} {\bibinfo {author} {\bibfnamefont {D.~A.}\ \bibnamefont
  {Wasson}},\ }\bibfield  {title} {\bibinfo {title} {{Comment on "Nuclear-bound
  quarkonium"}},\ }\href {https://doi.org/10.1103/PhysRevLett.67.2237}
  {\bibfield  {journal} {\bibinfo  {journal} {Phys. Rev. Lett.}\ }\textbf
  {\bibinfo {volume} {67}},\ \bibinfo {pages} {2237} (\bibinfo {year}
  {1991})}\BibitemShut {NoStop}%
\bibitem [{\citenamefont {Kim}\ and\ \citenamefont {Lee}(2001)}]{KIM2001517}%
  \BibitemOpen
  \bibfield  {author} {\bibinfo {author} {\bibfnamefont {S.}~\bibnamefont
  {Kim}}\ and\ \bibinfo {author} {\bibfnamefont {S.~H.}\ \bibnamefont {Lee}},\
  }\bibfield  {title} {\bibinfo {title} {{QCD sum rules for $J/\psi$ in the
  nuclear medium: calculation of the Wilson coefficients of gluon operators up
  to dimension 6}},\ }\href
  {https://doi.org/https://doi.org/10.1016/S0375-9474(00)00350-X} {\bibfield
  {journal} {\bibinfo  {journal} {Nucl. Phys. A}\ }\textbf {\bibinfo {volume}
  {679}},\ \bibinfo {pages} {517} (\bibinfo {year} {2001})}\BibitemShut
  {NoStop}%
\bibitem [{\citenamefont {Kumar}\ and\ \citenamefont
  {Mishra}(2010)}]{PhysRevC.82.045207}%
  \BibitemOpen
  \bibfield  {author} {\bibinfo {author} {\bibfnamefont {A.}~\bibnamefont
  {Kumar}}\ and\ \bibinfo {author} {\bibfnamefont {A.}~\bibnamefont {Mishra}},\
  }\bibfield  {title} {\bibinfo {title} {{$J/\ensuremath{\psi}$ and
  ${\ensuremath{\eta}}_{c}$ masses in isospin asymmetric hot nuclear matter: A
  QCD sum rule approach}},\ }\href {https://doi.org/10.1103/PhysRevC.82.045207}
  {\bibfield  {journal} {\bibinfo  {journal} {Phys. Rev. C}\ }\textbf {\bibinfo
  {volume} {82}},\ \bibinfo {pages} {045207} (\bibinfo {year}
  {2010})}\BibitemShut {NoStop}%
\bibitem [{\citenamefont {Sibirtsev}\ and\ \citenamefont
  {Voloshin}(2005)}]{PhysRevD.71.076005}%
  \BibitemOpen
  \bibfield  {author} {\bibinfo {author} {\bibfnamefont {A.}~\bibnamefont
  {Sibirtsev}}\ and\ \bibinfo {author} {\bibfnamefont {M.~B.}\ \bibnamefont
  {Voloshin}},\ }\bibfield  {title} {\bibinfo {title} {{Interaction of slow
  $J/\ensuremath{\psi}$ and ${\ensuremath{\psi}}^{\ensuremath{'}}$ with
  nucleons}},\ }\href {https://doi.org/10.1103/PhysRevD.71.076005} {\bibfield
  {journal} {\bibinfo  {journal} {Phys. Rev. D}\ }\textbf {\bibinfo {volume}
  {71}},\ \bibinfo {pages} {076005} (\bibinfo {year} {2005})}\BibitemShut
  {NoStop}%
\bibitem [{\citenamefont {Sibirtsev}\ \emph {et~al.}(2001)\citenamefont
  {Sibirtsev}, \citenamefont {Tsushima},\ and\ \citenamefont
  {Thomas}}]{PhysRevC.63.044906}%
  \BibitemOpen
  \bibfield  {author} {\bibinfo {author} {\bibfnamefont {A.}~\bibnamefont
  {Sibirtsev}}, \bibinfo {author} {\bibfnamefont {K.}~\bibnamefont
  {Tsushima}},\ and\ \bibinfo {author} {\bibfnamefont {A.~W.}\ \bibnamefont
  {Thomas}},\ }\bibfield  {title} {\bibinfo {title} {{Charmonium absorption by
  nucleons}},\ }\href {https://doi.org/10.1103/PhysRevC.63.044906} {\bibfield
  {journal} {\bibinfo  {journal} {Phys. Rev. C}\ }\textbf {\bibinfo {volume}
  {63}},\ \bibinfo {pages} {044906} (\bibinfo {year} {2001})}\BibitemShut
  {NoStop}%
\bibitem [{\citenamefont {Oh}\ \emph {et~al.}(2007)\citenamefont {Oh},
  \citenamefont {Liu},\ and\ \citenamefont {Ko}}]{PhysRevC.75.064903}%
  \BibitemOpen
  \bibfield  {author} {\bibinfo {author} {\bibfnamefont {Y.}~\bibnamefont
  {Oh}}, \bibinfo {author} {\bibfnamefont {W.}~\bibnamefont {Liu}},\ and\
  \bibinfo {author} {\bibfnamefont {C.~M.}\ \bibnamefont {Ko}},\ }\bibfield
  {title} {\bibinfo {title} {{$J/\ensuremath{\psi}$ absorption by nucleons in
  the meson-exchange model}},\ }\href
  {https://doi.org/10.1103/PhysRevC.75.064903} {\bibfield  {journal} {\bibinfo
  {journal} {Phys. Rev. C}\ }\textbf {\bibinfo {volume} {75}},\ \bibinfo
  {pages} {064903} (\bibinfo {year} {2007})}\BibitemShut {NoStop}%
\bibitem [{\citenamefont {Hilbert}\ \emph {et~al.}(2007)\citenamefont
  {Hilbert}, \citenamefont {Black}, \citenamefont {Barnes},\ and\ \citenamefont
  {Swanson}}]{PhysRevC.75.064907}%
  \BibitemOpen
  \bibfield  {author} {\bibinfo {author} {\bibfnamefont {J.~P.}\ \bibnamefont
  {Hilbert}}, \bibinfo {author} {\bibfnamefont {N.}~\bibnamefont {Black}},
  \bibinfo {author} {\bibfnamefont {T.}~\bibnamefont {Barnes}},\ and\ \bibinfo
  {author} {\bibfnamefont {E.~S.}\ \bibnamefont {Swanson}},\ }\bibfield
  {title} {\bibinfo {title} {{Charmonium-nucleon dissociation cross sections in
  the quark model}},\ }\href {https://doi.org/10.1103/PhysRevC.75.064907}
  {\bibfield  {journal} {\bibinfo  {journal} {Phys. Rev. C}\ }\textbf {\bibinfo
  {volume} {75}},\ \bibinfo {pages} {064907} (\bibinfo {year}
  {2007})}\BibitemShut {NoStop}%
\bibitem [{\citenamefont {Etminan}(2025{\natexlab{a}})}]{ETMINAN2025PLB139564}%
  \BibitemOpen
  \bibfield  {author} {\bibinfo {author} {\bibfnamefont {F.}~\bibnamefont
  {Etminan}},\ }\bibfield  {title} {\bibinfo {title} {{Exploring the $
  \phi-\alpha $ interaction via femtoscopic study}},\ }\href
  {https://doi.org/https://doi.org/10.1016/j.physletb.2025.139564} {\bibfield
  {journal} {\bibinfo  {journal} {Phys. Lett. B}\ }\textbf {\bibinfo {volume}
  {866}},\ \bibinfo {pages} {139564} (\bibinfo {year}
  {2025}{\natexlab{a}})}\BibitemShut {NoStop}%
\bibitem [{\citenamefont {Etminan}(2026)}]{etminan2026SR}%
  \BibitemOpen
  \bibfield  {author} {\bibinfo {author} {\bibfnamefont {F.}~\bibnamefont
  {Etminan}},\ }\bibfield  {title} {\bibinfo {title} {Quantitative predictions
  of alpha-charmonium correlation functions in high-energy collisions},\ }\href
  {https://doi.org/10.1038/s41598-026-51720-7} {\bibfield  {journal} {\bibinfo
  {journal} {Sci. Rep.}\ }\textbf {\bibinfo {volume} {16}},\ \bibinfo {pages}
  {20811} (\bibinfo {year} {2026})}\BibitemShut {NoStop}%
\bibitem [{\citenamefont {de~T\'eramond}\ \emph {et~al.}(1998)\citenamefont
  {de~T\'eramond}, \citenamefont {Espinoza},\ and\ \citenamefont
  {Ortega-Rodr\'{\i}guez}}]{PhysRevD.58.034012}%
  \BibitemOpen
  \bibfield  {author} {\bibinfo {author} {\bibfnamefont {G.~F.}\ \bibnamefont
  {de~T\'eramond}}, \bibinfo {author} {\bibfnamefont {R.}~\bibnamefont
  {Espinoza}},\ and\ \bibinfo {author} {\bibfnamefont {M.}~\bibnamefont
  {Ortega-Rodr\'{\i}guez}},\ }\bibfield  {title} {\bibinfo {title}
  {{Proton-proton spin correlations at charm threshold and quarkonium bound to
  nuclei}},\ }\href {https://doi.org/10.1103/PhysRevD.58.034012} {\bibfield
  {journal} {\bibinfo  {journal} {Phys. Rev. D}\ }\textbf {\bibinfo {volume}
  {58}},\ \bibinfo {pages} {034012} (\bibinfo {year} {1998})}\BibitemShut
  {NoStop}%
\bibitem [{\citenamefont {Cobos-Martínez}\ \emph {et~al.}(2020)\citenamefont
  {Cobos-Martínez}, \citenamefont {Tsushima}, \citenamefont {Krein},\ and\
  \citenamefont {Thomas}}]{COBOSMARTINEZ2020135882}%
  \BibitemOpen
  \bibfield  {author} {\bibinfo {author} {\bibfnamefont {J.}~\bibnamefont
  {Cobos-Martínez}}, \bibinfo {author} {\bibfnamefont {K.}~\bibnamefont
  {Tsushima}}, \bibinfo {author} {\bibfnamefont {G.}~\bibnamefont {Krein}},\
  and\ \bibinfo {author} {\bibfnamefont {A.}~\bibnamefont {Thomas}},\
  }\bibfield  {title} {\bibinfo {title} {{$ \eta_{c}$-nucleus bound states}},\
  }\href {https://doi.org/https://doi.org/10.1016/j.physletb.2020.135882}
  {\bibfield  {journal} {\bibinfo  {journal} {Phys. Lett. B}\ }\textbf
  {\bibinfo {volume} {811}},\ \bibinfo {pages} {135882} (\bibinfo {year}
  {2020})}\BibitemShut {NoStop}%
\bibitem [{\citenamefont {Ishii}\ \emph {et~al.}(2012)\citenamefont {Ishii}
  \emph {et~al.}}]{ishii2012}%
  \BibitemOpen
  \bibfield  {author} {\bibinfo {author} {\bibfnamefont {N.}~\bibnamefont
  {Ishii}} \emph {et~al.} (\bibinfo {collaboration} {HAL QCD Collaboration}),\
  }\bibfield  {title} {\bibinfo {title} {{Hadron–hadron interactions from
  imaginary-time Nambu–Bethe–Salpeter wave function on the lattice}},\
  }\href {https://doi.org/https://doi.org/10.1016/j.physletb.2012.04.076}
  {\bibfield  {journal} {\bibinfo  {journal} {Phys. Lett. B}\ }\textbf
  {\bibinfo {volume} {712}},\ \bibinfo {pages} {437} (\bibinfo {year}
  {2012})}\BibitemShut {NoStop}%
\bibitem [{\citenamefont {Aoki}\ \emph {et~al.}(2013)\citenamefont {Aoki},
  \citenamefont {Charron}, \citenamefont {Doi}, \citenamefont {Hatsuda},
  \citenamefont {Inoue},\ and\ \citenamefont {Ishii}}]{aoki2013}%
  \BibitemOpen
  \bibfield  {author} {\bibinfo {author} {\bibfnamefont {S.}~\bibnamefont
  {Aoki}}, \bibinfo {author} {\bibfnamefont {B.}~\bibnamefont {Charron}},
  \bibinfo {author} {\bibfnamefont {T.}~\bibnamefont {Doi}}, \bibinfo {author}
  {\bibfnamefont {T.}~\bibnamefont {Hatsuda}}, \bibinfo {author} {\bibfnamefont
  {T.}~\bibnamefont {Inoue}},\ and\ \bibinfo {author} {\bibfnamefont
  {N.}~\bibnamefont {Ishii}},\ }\bibfield  {title} {\bibinfo {title}
  {Construction of energy-independent potentials above inelastic thresholds in
  quantum field theories},\ }\href {https://doi.org/10.1103/PhysRevD.87.034512}
  {\bibfield  {journal} {\bibinfo  {journal} {Phys. Rev. D}\ }\textbf {\bibinfo
  {volume} {87}},\ \bibinfo {pages} {034512} (\bibinfo {year}
  {2013})}\BibitemShut {NoStop}%
\bibitem [{\citenamefont {Iritani}\ \emph {et~al.}(2019)\citenamefont {Iritani}
  \emph {et~al.}}]{Iritani2019prb}%
  \BibitemOpen
  \bibfield  {author} {\bibinfo {author} {\bibfnamefont {T.}~\bibnamefont
  {Iritani}} \emph {et~al.},\ }\bibfield  {title} {\bibinfo {title} {$
  \mathrm{N} \mathrm{\ensuremath{\Omega}} $ dibaryon from lattice $
  \mathrm{QCD}$ near the physical point},\ }\href
  {https://doi.org/https://doi.org/10.1016/j.physletb.2019.03.050} {\bibfield
  {journal} {\bibinfo  {journal} {Phys. Lett. B}\ }\textbf {\bibinfo {volume}
  {792}},\ \bibinfo {pages} {284} (\bibinfo {year} {2019})}\BibitemShut
  {NoStop}%
\bibitem [{\citenamefont {Kenji}\ \emph {et~al.}(2020)\citenamefont {Kenji}
  \emph {et~al.}}]{sasaki2020}%
  \BibitemOpen
  \bibfield  {author} {\bibinfo {author} {\bibfnamefont {S.}~\bibnamefont
  {Kenji}} \emph {et~al.} (\bibinfo {collaboration} {HAL QCD Collaboration}),\
  }\bibfield  {title} {\bibinfo {title} {$
  \mathrm{\ensuremath{\Lambda}}\mathrm{\ensuremath{\Lambda}} $ and $\mathrm{N}
  \mathrm{\ensuremath{\Xi}}$ interactions from lattice $\mathrm{QCD}$ near the
  physical point},\ }\href
  {https://doi.org/https://doi.org/10.1016/j.nuclphysa.2020.121737} {\bibfield
  {journal} {\bibinfo  {journal} {Nucl. Phys. A}\ }\textbf {\bibinfo {volume}
  {998}},\ \bibinfo {pages} {121737} (\bibinfo {year} {2020})}\BibitemShut
  {NoStop}%
\bibitem [{\citenamefont {Etminan}\ \emph {et~al.}(2024)\citenamefont
  {Etminan}, \citenamefont {Sasaki},\ and\ \citenamefont
  {Inoue}}]{etminan2024prd}%
  \BibitemOpen
  \bibfield  {author} {\bibinfo {author} {\bibfnamefont {F.}~\bibnamefont
  {Etminan}}, \bibinfo {author} {\bibfnamefont {K.}~\bibnamefont {Sasaki}},\
  and\ \bibinfo {author} {\bibfnamefont {T.}~\bibnamefont {Inoue}},\ }\bibfield
   {title} {\bibinfo {title} {{Coupled-channel
  ${\mathrm{\ensuremath{\Lambda}}}_{c}{K}^{+}\ensuremath{-}p{D}_{s}$
  interaction in the flavor SU(3) limit of lattice QCD}},\ }\href
  {https://doi.org/10.1103/PhysRevD.109.074506} {\bibfield  {journal} {\bibinfo
   {journal} {Phys. Rev. D}\ }\textbf {\bibinfo {volume} {109}},\ \bibinfo
  {pages} {074506} (\bibinfo {year} {2024})}\BibitemShut {NoStop}%
\bibitem [{\citenamefont {Lyu}\ \emph {et~al.}(2025)\citenamefont {Lyu},
  \citenamefont {Doi}, \citenamefont {Hatsuda},\ and\ \citenamefont
  {Sugiura}}]{LyuPLB2025}%
  \BibitemOpen
  \bibfield  {author} {\bibinfo {author} {\bibfnamefont {Y.}~\bibnamefont
  {Lyu}}, \bibinfo {author} {\bibfnamefont {T.}~\bibnamefont {Doi}}, \bibinfo
  {author} {\bibfnamefont {T.}~\bibnamefont {Hatsuda}},\ and\ \bibinfo {author}
  {\bibfnamefont {T.}~\bibnamefont {Sugiura}},\ }\bibfield  {title} {\bibinfo
  {title} {{Nucleon-charmonium interactions from lattice QCD}},\ }\href
  {https://doi.org/https://doi.org/10.1016/j.physletb.2024.139178} {\bibfield
  {journal} {\bibinfo  {journal} {Phys. Lett. B}\ }\textbf {\bibinfo {volume}
  {860}},\ \bibinfo {pages} {139178} (\bibinfo {year} {2025})}\BibitemShut
  {NoStop}%
\bibitem [{\citenamefont {Beane}\ \emph {et~al.}(2015)\citenamefont {Beane},
  \citenamefont {Chang}, \citenamefont {Cohen}, \citenamefont {Detmold},
  \citenamefont {Lin}, \citenamefont {Orginos}, \citenamefont {Parre\~no},\
  and\ \citenamefont {Savage}}]{Beane2015}%
  \BibitemOpen
  \bibfield  {author} {\bibinfo {author} {\bibfnamefont {S.~R.}\ \bibnamefont
  {Beane}}, \bibinfo {author} {\bibfnamefont {E.}~\bibnamefont {Chang}},
  \bibinfo {author} {\bibfnamefont {S.~D.}\ \bibnamefont {Cohen}}, \bibinfo
  {author} {\bibfnamefont {W.}~\bibnamefont {Detmold}}, \bibinfo {author}
  {\bibfnamefont {H.-W.}\ \bibnamefont {Lin}}, \bibinfo {author} {\bibfnamefont
  {K.}~\bibnamefont {Orginos}}, \bibinfo {author} {\bibfnamefont
  {A.}~\bibnamefont {Parre\~no}},\ and\ \bibinfo {author} {\bibfnamefont
  {M.~J.}\ \bibnamefont {Savage}} (\bibinfo {collaboration} {NPLQCD
  Collaboration}),\ }\bibfield  {title} {\bibinfo {title} {{Quarkonium-nucleus
  bound states from lattice QCD}},\ }\href
  {https://doi.org/10.1103/PhysRevD.91.114503} {\bibfield  {journal} {\bibinfo
  {journal} {Phys. Rev. D}\ }\textbf {\bibinfo {volume} {91}},\ \bibinfo
  {pages} {114503} (\bibinfo {year} {2015})}\BibitemShut {NoStop}%
\bibitem [{\citenamefont
  {Etminan}(2025{\natexlab{b}})}]{EtminanPRCalphaAlphaccbar2025}%
  \BibitemOpen
  \bibfield  {author} {\bibinfo {author} {\bibfnamefont {F.}~\bibnamefont
  {Etminan}},\ }\bibfield  {title} {\bibinfo {title} {{Bound states of
  $_{c\overline{c}}^{9}\mathrm{Be}$ within
  $c\overline{c}+\ensuremath{\alpha}+\ensuremath{\alpha}$ cluster models based
  on state-of-the-art QCD charmonium-nucleon interactions}},\ }\href
  {https://doi.org/10.1103/wwjr-3x9l} {\bibfield  {journal} {\bibinfo
  {journal} {Phys. Rev. C}\ }\textbf {\bibinfo {volume} {112}},\ \bibinfo
  {pages} {064001} (\bibinfo {year} {2025}{\natexlab{b}})}\BibitemShut
  {NoStop}%
\bibitem [{\citenamefont {Thompson}\ and\ \citenamefont
  {Zhukov}(1996)}]{ThompsonPRC1996}%
  \BibitemOpen
  \bibfield  {author} {\bibinfo {author} {\bibfnamefont {I.~J.}\ \bibnamefont
  {Thompson}}\ and\ \bibinfo {author} {\bibfnamefont {M.~V.}\ \bibnamefont
  {Zhukov}},\ }\bibfield  {title} {\bibinfo {title} {Structure and reactions of
  the $^{12,14}\mathrm{Be}$ nuclei},\ }\href
  {https://doi.org/10.1103/PhysRevC.53.708} {\bibfield  {journal} {\bibinfo
  {journal} {Phys. Rev. C}\ }\textbf {\bibinfo {volume} {53}},\ \bibinfo
  {pages} {708} (\bibinfo {year} {1996})}\BibitemShut {NoStop}%
\bibitem [{\citenamefont {Thompson}\ \emph {et~al.}(2004)\citenamefont
  {Thompson}, \citenamefont {Nunes},\ and\ \citenamefont {Danilin}}]{face}%
  \BibitemOpen
  \bibfield  {author} {\bibinfo {author} {\bibfnamefont {I.}~\bibnamefont
  {Thompson}}, \bibinfo {author} {\bibfnamefont {F.}~\bibnamefont {Nunes}},\
  and\ \bibinfo {author} {\bibfnamefont {B.}~\bibnamefont {Danilin}},\
  }\bibfield  {title} {\bibinfo {title} {$\mathrm{FaCE}$: a tool for three body
  $\mathrm{Faddeev}$ calculations with core excitation},\ }\href
  {https://doi.org/https://doi.org/10.1016/j.cpc.2004.03.007} {\bibfield
  {journal} {\bibinfo  {journal} {Comput. Phys. Commun}\ }\textbf {\bibinfo
  {volume} {161}},\ \bibinfo {pages} {87} (\bibinfo {year} {2004})}\BibitemShut
  {NoStop}%
\bibitem [{\citenamefont {Zhukov}\ \emph {et~al.}(1993)\citenamefont {Zhukov}
  \emph {et~al.}}]{Zhukov93}%
  \BibitemOpen
  \bibfield  {author} {\bibinfo {author} {\bibfnamefont {M.}~\bibnamefont
  {Zhukov}} \emph {et~al.},\ }\bibfield  {title} {\bibinfo {title} {{Bound
  state properties of Borromean halo nuclei: $ ^{6}\mathrm{He} $ and $
  ^{11}\mathrm{Li} $}},\ }\href
  {https://doi.org/https://doi.org/10.1016/0370-1573(93)90141-Y} {\bibfield
  {journal} {\bibinfo  {journal} {Phys. Rep.}\ }\textbf {\bibinfo {volume}
  {231}},\ \bibinfo {pages} {151} (\bibinfo {year} {1993})}\BibitemShut
  {NoStop}%
\bibitem [{\citenamefont {Raynal}\ and\ \citenamefont
  {Revai}(1970)}]{raynal1970}%
  \BibitemOpen
  \bibfield  {author} {\bibinfo {author} {\bibfnamefont {J.}~\bibnamefont
  {Raynal}}\ and\ \bibinfo {author} {\bibfnamefont {J.}~\bibnamefont {Revai}},\
  }\bibfield  {title} {\bibinfo {title} {Transformation coefficients in the
  hyperspherical approach to the three-body problem},\ }\href
  {https://doi.org/10.1007/BF02756127} {\bibfield  {journal} {\bibinfo
  {journal} {Il Nuovo Cimento A (1965-1970)}\ }\textbf {\bibinfo {volume}
  {68}},\ \bibinfo {pages} {612} (\bibinfo {year} {1970})}\BibitemShut
  {NoStop}%
\bibitem [{\citenamefont {Etminan}\ and\ \citenamefont
  {Aalimi}(2024)}]{etminan2024prc}%
  \BibitemOpen
  \bibfield  {author} {\bibinfo {author} {\bibfnamefont {F.}~\bibnamefont
  {Etminan}}\ and\ \bibinfo {author} {\bibfnamefont {A.}~\bibnamefont
  {Aalimi}},\ }\bibfield  {title} {\bibinfo {title} {{Examination of the
  $\ensuremath{\phi}\text{\ensuremath{-}}NN$ bound-state problem with lattice
  QCD $N\text{\ensuremath{-}}\ensuremath{\phi}$ potentials}},\ }\href
  {https://doi.org/10.1103/PhysRevC.109.054002} {\bibfield  {journal} {\bibinfo
   {journal} {Phys. Rev. C}\ }\textbf {\bibinfo {volume} {109}},\ \bibinfo
  {pages} {054002} (\bibinfo {year} {2024})}\BibitemShut {NoStop}%
\bibitem [{\citenamefont {Casal}\ \emph {et~al.}(2020)\citenamefont {Casal},
  \citenamefont {Singh}, \citenamefont {Fortunato}, \citenamefont {Horiuchi},\
  and\ \citenamefont {Vitturi}}]{Casal2020}%
  \BibitemOpen
  \bibfield  {author} {\bibinfo {author} {\bibfnamefont {J.}~\bibnamefont
  {Casal}}, \bibinfo {author} {\bibfnamefont {J.}~\bibnamefont {Singh}},
  \bibinfo {author} {\bibfnamefont {L.}~\bibnamefont {Fortunato}}, \bibinfo
  {author} {\bibfnamefont {W.}~\bibnamefont {Horiuchi}},\ and\ \bibinfo
  {author} {\bibfnamefont {A.}~\bibnamefont {Vitturi}},\ }\bibfield  {title}
  {\bibinfo {title} {Electric dipole response of low-lying excitations in the
  two-neutron halo nucleus $^{29}\mathrm{F}$},\ }\href
  {https://doi.org/10.1103/PhysRevC.102.064627} {\bibfield  {journal} {\bibinfo
   {journal} {Phys. Rev. C}\ }\textbf {\bibinfo {volume} {102}},\ \bibinfo
  {pages} {064627} (\bibinfo {year} {2020})}\BibitemShut {NoStop}%
\bibitem [{\citenamefont {Tanihata}\ \emph {et~al.}(1988)\citenamefont
  {Tanihata} \emph {et~al.}}]{TANIHATA1988592}%
  \BibitemOpen
  \bibfield  {author} {\bibinfo {author} {\bibfnamefont {I.}~\bibnamefont
  {Tanihata}} \emph {et~al.},\ }\bibfield  {title} {\bibinfo {title}
  {{Measurement of interaction cross sections using isotope beams of Be and B
  and isospin dependence of the nuclear radii}},\ }\href
  {https://doi.org/https://doi.org/10.1016/0370-2693(88)90702-2} {\bibfield
  {journal} {\bibinfo  {journal} {Phys. Lett. B}\ }\textbf {\bibinfo {volume}
  {206}},\ \bibinfo {pages} {592} (\bibinfo {year} {1988})}\BibitemShut
  {NoStop}%
\bibitem [{\citenamefont {Povh}\ and\ \citenamefont
  {Hüfner}(1990)}]{POVH1990653}%
  \BibitemOpen
  \bibfield  {author} {\bibinfo {author} {\bibfnamefont {B.}~\bibnamefont
  {Povh}}\ and\ \bibinfo {author} {\bibfnamefont {J.}~\bibnamefont {Hüfner}},\
  }\bibfield  {title} {\bibinfo {title} {{Systematics of strong interaction
  radii for hadrons}},\ }\href
  {https://doi.org/https://doi.org/10.1016/0370-2693(90)90707-D} {\bibfield
  {journal} {\bibinfo  {journal} {Phys. Lett. B}\ }\textbf {\bibinfo {volume}
  {245}},\ \bibinfo {pages} {653} (\bibinfo {year} {1990})}\BibitemShut
  {NoStop}%
\bibitem [{\citenamefont {Brambilla}\ \emph {et~al.}(2016)\citenamefont
  {Brambilla}, \citenamefont {Krein}, \citenamefont {Tarr\'us~Castell\`a},\
  and\ \citenamefont {Vairo}}]{Brambilla2016}%
  \BibitemOpen
  \bibfield  {author} {\bibinfo {author} {\bibfnamefont {N.}~\bibnamefont
  {Brambilla}}, \bibinfo {author} {\bibfnamefont {G.~a.}\ \bibnamefont
  {Krein}}, \bibinfo {author} {\bibfnamefont {J.}~\bibnamefont
  {Tarr\'us~Castell\`a}},\ and\ \bibinfo {author} {\bibfnamefont
  {A.}~\bibnamefont {Vairo}},\ }\bibfield  {title} {\bibinfo {title}
  {{Long-range properties of $1S$ bottomonium states}},\ }\href
  {https://doi.org/10.1103/PhysRevD.93.054002} {\bibfield  {journal} {\bibinfo
  {journal} {Phys. Rev. D}\ }\textbf {\bibinfo {volume} {93}},\ \bibinfo
  {pages} {054002} (\bibinfo {year} {2016})}\BibitemShut {NoStop}%
\bibitem [{\citenamefont {Tarr\'us~Castell\`a}\ and\ \citenamefont
  {Krein}(2018)}]{PhysRevD.98.014029}%
  \BibitemOpen
  \bibfield  {author} {\bibinfo {author} {\bibfnamefont {J.}~\bibnamefont
  {Tarr\'us~Castell\`a}}\ and\ \bibinfo {author} {\bibfnamefont {G.~a.}\
  \bibnamefont {Krein}},\ }\bibfield  {title} {\bibinfo {title} {{Effective
  field theory for the nucleon-quarkonium interaction}},\ }\href
  {https://doi.org/10.1103/PhysRevD.98.014029} {\bibfield  {journal} {\bibinfo
  {journal} {Phys. Rev. D}\ }\textbf {\bibinfo {volume} {98}},\ \bibinfo
  {pages} {014029} (\bibinfo {year} {2018})}\BibitemShut {NoStop}%
\bibitem [{\citenamefont {Satchler}\ and\ \citenamefont
  {Love}(1979)}]{Satchler1979}%
  \BibitemOpen
  \bibfield  {author} {\bibinfo {author} {\bibfnamefont {G.}~\bibnamefont
  {Satchler}}\ and\ \bibinfo {author} {\bibfnamefont {W.}~\bibnamefont
  {Love}},\ }\bibfield  {title} {\bibinfo {title} {Folding model potentials
  from realistic interactions for heavy-ion scattering},\ }\href
  {https://doi.org/https://doi.org/10.1016/0370-1573(79)90081-4} {\bibfield
  {journal} {\bibinfo  {journal} {Phys. Rep.}\ }\textbf {\bibinfo {volume}
  {55}},\ \bibinfo {pages} {183} (\bibinfo {year} {1979})}\BibitemShut
  {NoStop}%
\bibitem [{\citenamefont {Etminan}\ and\ \citenamefont
  {Firoozabadi}(2020)}]{Etminan:2019gds}%
  \BibitemOpen
  \bibfield  {author} {\bibinfo {author} {\bibfnamefont {F.}~\bibnamefont
  {Etminan}}\ and\ \bibinfo {author} {\bibfnamefont {M.~M.}\ \bibnamefont
  {Firoozabadi}},\ }\bibfield  {title} {\bibinfo {title} {{Simple Woods-Saxon
  type form for $\Omega \alpha $ and $\Xi \alpha$ interactions using folding
  model}},\ }\href {https://doi.org/10.1088/1674-1137/44/5/054106} {\bibfield
  {journal} {\bibinfo  {journal} {Chin. Phys. C}\ }\textbf {\bibinfo {volume}
  {44}},\ \bibinfo {pages} {054106} (\bibinfo {year} {2020})}\BibitemShut
  {NoStop}%
\bibitem [{\citenamefont {Gross}(1993)}]{Lenz1993}%
  \BibitemOpen
  \bibfield  {author} {\bibinfo {author} {\bibfnamefont {D.}~\bibnamefont
  {Gross}},\ }\bibinfo {title} {The friction model for deep-inelastic and
  fusion reactions},\ in\ \href {https://doi.org/10.1007/978-1-4613-9335-1_11}
  {\emph {\bibinfo {booktitle} {Computational Nuclear Physics 2: Nuclear
  Reactions}}},\ \bibinfo {editor} {edited by\ \bibinfo {editor} {\bibfnamefont
  {K.}~\bibnamefont {Langanke}}, \bibinfo {editor} {\bibfnamefont {J.~A.}\
  \bibnamefont {Maruhn}},\ and\ \bibinfo {editor} {\bibfnamefont {S.~E.}\
  \bibnamefont {Koonin}}}\ (\bibinfo  {publisher} {Springer New York},\
  \bibinfo {address} {New York, NY},\ \bibinfo {year} {1993})\ pp.\ \bibinfo
  {pages} {148--154}\BibitemShut {NoStop}%
\bibitem [{\citenamefont {{Joint Committee for Guides in
  Metrology}}(2008)}]{JCGM101}%
  \BibitemOpen
  \bibfield  {author} {\bibinfo {author} {\bibnamefont {{Joint Committee for
  Guides in Metrology}}},\ }\href@noop {} {\bibinfo {title} {{Evaluation of
  Measurement Data -- Supplement 1 to the Guide to the Expression of
  Uncertainty in Measurement: Propagation of Distributions Using a Monte Carlo
  Method}}} (\bibinfo {year} {2008}),\ \bibinfo {note} {{JCGM
  101:2008}}\BibitemShut {NoStop}%
\bibitem [{\citenamefont {Filikhin}\ \emph {et~al.}(2005)\citenamefont
  {Filikhin}, \citenamefont {Suslov},\ and\ \citenamefont
  {Vlahovic}}]{Filikhin_2005}%
  \BibitemOpen
  \bibfield  {author} {\bibinfo {author} {\bibfnamefont {I.}~\bibnamefont
  {Filikhin}}, \bibinfo {author} {\bibfnamefont {V.~M.}\ \bibnamefont
  {Suslov}},\ and\ \bibinfo {author} {\bibfnamefont {B.}~\bibnamefont
  {Vlahovic}},\ }\bibfield  {title} {\bibinfo {title} {{ A new prediction for
  the binding energy of the $_{\Lambda}^{7}$He hypernucleus}},\ }\href
  {https://doi.org/10.1088/0954-3899/31/5/009} {\bibfield  {journal} {\bibinfo
  {journal} {Phys. G: Nucl. Part. Phys.}\ }\textbf {\bibinfo {volume} {31}},\
  \bibinfo {pages} {389} (\bibinfo {year} {2005})}\BibitemShut {NoStop}%
\bibitem [{\citenamefont {Hiyama}\ \emph {et~al.}(2022)\citenamefont {Hiyama},
  \citenamefont {Isaka}, \citenamefont {Doi},\ and\ \citenamefont
  {Hatsuda}}]{HiyamaPRC2022}%
  \BibitemOpen
  \bibfield  {author} {\bibinfo {author} {\bibfnamefont {E.}~\bibnamefont
  {Hiyama}}, \bibinfo {author} {\bibfnamefont {M.}~\bibnamefont {Isaka}},
  \bibinfo {author} {\bibfnamefont {T.}~\bibnamefont {Doi}},\ and\ \bibinfo
  {author} {\bibfnamefont {T.}~\bibnamefont {Hatsuda}},\ }\bibfield  {title}
  {\bibinfo {title} {{Probing the $\mathrm{\ensuremath{\Xi}}N$ interaction
  through inversion of spin-doublets in
  $\mathrm{\ensuremath{\Xi}}N\ensuremath{\alpha}\ensuremath{\alpha}$ nuclei}},\
  }\href {https://doi.org/10.1103/PhysRevC.106.064318} {\bibfield  {journal}
  {\bibinfo  {journal} {Phys. Rev. C}\ }\textbf {\bibinfo {volume} {106}},\
  \bibinfo {pages} {064318} (\bibinfo {year} {2022})}\BibitemShut {NoStop}%
\bibitem [{\citenamefont {Filikhin}\ \emph {et~al.}(2024)\citenamefont
  {Filikhin}, \citenamefont {Kezerashvili},\ and\ \citenamefont
  {Vlahovic}}]{filikhin2024phihe}%
  \BibitemOpen
  \bibfield  {author} {\bibinfo {author} {\bibfnamefont {I.}~\bibnamefont
  {Filikhin}}, \bibinfo {author} {\bibfnamefont {R.~Y.}\ \bibnamefont
  {Kezerashvili}},\ and\ \bibinfo {author} {\bibfnamefont {B.}~\bibnamefont
  {Vlahovic}},\ }\bibfield  {title} {\bibinfo {title} {{Bound states of
  $_{\ensuremath{\phi}}^{9}\mathrm{Be}$ and
  $_{\ensuremath{\phi}\ensuremath{\phi}}^{6}\mathrm{He}$ within
  $\ensuremath{\phi}+\ensuremath{\alpha}+\ensuremath{\alpha}$ and
  $\ensuremath{\phi}+\ensuremath{\phi}+\ensuremath{\alpha}$ cluster models}},\
  }\href {https://doi.org/10.1103/PhysRevC.110.065202} {\bibfield  {journal}
  {\bibinfo  {journal} {Phys. Rev. C}\ }\textbf {\bibinfo {volume} {110}},\
  \bibinfo {pages} {065202} (\bibinfo {year} {2024})}\BibitemShut {NoStop}%
\bibitem [{\citenamefont
  {Etminan}(2025{\natexlab{c}})}]{etminanPRCphiNalpha2025}%
  \BibitemOpen
  \bibfield  {author} {\bibinfo {author} {\bibfnamefont {F.}~\bibnamefont
  {Etminan}},\ }\bibfield  {title} {\bibinfo {title} {{Probing
  $\ensuremath{\phi}N$ interaction through bound states of the
  $\ensuremath{\phi}N\text{\ensuremath{-}}\ensuremath{\alpha}$ system}},\
  }\href {https://doi.org/10.1103/p93c-zk34} {\bibfield  {journal} {\bibinfo
  {journal} {Phys. Rev. C}\ }\textbf {\bibinfo {volume} {112}},\ \bibinfo
  {pages} {044003} (\bibinfo {year} {2025}{\natexlab{c}})}\BibitemShut
  {NoStop}%
\bibitem [{\citenamefont {Filikhin}(2000)}]{filikhin2000alpha8be}%
  \BibitemOpen
  \bibfield  {author} {\bibinfo {author} {\bibfnamefont {I.}~\bibnamefont
  {Filikhin}},\ }\bibfield  {title} {\bibinfo {title} {{$\alpha$ $^8$Be cluster
  model for the $ 0_{2}^{+} $ resonance in the $^{12}$C nucleus}},\ }\href
  {https://doi.org/10.1134/1.1312886} {\bibfield  {journal} {\bibinfo
  {journal} {Phys. At. Nucl.}\ }\textbf {\bibinfo {volume} {63}},\ \bibinfo
  {pages} {1527} (\bibinfo {year} {2000})}\BibitemShut {NoStop}%
\bibitem [{\citenamefont {Nunes}\ \emph {et~al.}(1996)\citenamefont {Nunes},
  \citenamefont {Christley}, \citenamefont {Thompson}, \citenamefont
  {Johnson},\ and\ \citenamefont {Efros}}]{NUNES199643}%
  \BibitemOpen
  \bibfield  {author} {\bibinfo {author} {\bibfnamefont {F.}~\bibnamefont
  {Nunes}}, \bibinfo {author} {\bibfnamefont {J.}~\bibnamefont {Christley}},
  \bibinfo {author} {\bibfnamefont {I.}~\bibnamefont {Thompson}}, \bibinfo
  {author} {\bibfnamefont {R.}~\bibnamefont {Johnson}},\ and\ \bibinfo {author}
  {\bibfnamefont {V.}~\bibnamefont {Efros}},\ }\bibfield  {title} {\bibinfo
  {title} {{Core excitation in three-body systems: Application to
  $^{12}\textrm{Be}$}},\ }\href
  {https://doi.org/https://doi.org/10.1016/0375-9474(96)00284-9} {\bibfield
  {journal} {\bibinfo  {journal} {Nucl. Phys. A}\ }\textbf {\bibinfo {volume}
  {609}},\ \bibinfo {pages} {43} (\bibinfo {year} {1996})}\BibitemShut
  {NoStop}%
\bibitem [{\citenamefont {Nunes}\ \emph {et~al.}(2002)\citenamefont {Nunes},
  \citenamefont {Thompson},\ and\ \citenamefont {Tostevin}}]{NUNES2002593}%
  \BibitemOpen
  \bibfield  {author} {\bibinfo {author} {\bibfnamefont {F.}~\bibnamefont
  {Nunes}}, \bibinfo {author} {\bibfnamefont {I.}~\bibnamefont {Thompson}},\
  and\ \bibinfo {author} {\bibfnamefont {J.}~\bibnamefont {Tostevin}},\
  }\bibfield  {title} {\bibinfo {title} {{Core excitation in
  $^{12}\textrm{Be}$}},\ }\href
  {https://doi.org/https://doi.org/10.1016/S0375-9474(01)01667-0} {\bibfield
  {journal} {\bibinfo  {journal} {Nucl. Phys. A}\ }\textbf {\bibinfo {volume}
  {703}},\ \bibinfo {pages} {593} (\bibinfo {year} {2002})}\BibitemShut
  {NoStop}%
\bibitem [{\citenamefont {Audi}\ and\ \citenamefont
  {Wapstra}(1993)}]{AUDI199366}%
  \BibitemOpen
  \bibfield  {author} {\bibinfo {author} {\bibfnamefont {G.}~\bibnamefont
  {Audi}}\ and\ \bibinfo {author} {\bibfnamefont {A.}~\bibnamefont {Wapstra}},\
  }\bibfield  {title} {\bibinfo {title} {{The 1993 atomic mass evaluation: (II)
  Nuclear-reaction and separation energies}},\ }\href
  {https://doi.org/https://doi.org/10.1016/0375-9474(93)90025-S} {\bibfield
  {journal} {\bibinfo  {journal} {Nucl. Phys. A}\ }\textbf {\bibinfo {volume}
  {565}},\ \bibinfo {pages} {66} (\bibinfo {year} {1993})}\BibitemShut
  {NoStop}%
\bibitem [{\citenamefont {Palit}\ \emph {et~al.}(2003)\citenamefont {Palit}
  \emph {et~al.}}]{PhysRevC.68.034318}%
  \BibitemOpen
  \bibfield  {author} {\bibinfo {author} {\bibfnamefont {R.}~\bibnamefont
  {Palit}} \emph {et~al.} (\bibinfo {collaboration} {LAND/FRS Collaboration}),\
  }\bibfield  {title} {\bibinfo {title} {{Exclusive measurement of breakup
  reactions with the one-neutron halo nucleus ${}^{11}\mathrm{Be}$}},\ }\href
  {https://doi.org/10.1103/PhysRevC.68.034318} {\bibfield  {journal} {\bibinfo
  {journal} {Phys. Rev. C}\ }\textbf {\bibinfo {volume} {68}},\ \bibinfo
  {pages} {034318} (\bibinfo {year} {2003})}\BibitemShut {NoStop}%
\bibitem [{\citenamefont {Al-Khalili}\ and\ \citenamefont
  {Tostevin}(1996)}]{PhysRevLett.76.3903}%
  \BibitemOpen
  \bibfield  {author} {\bibinfo {author} {\bibfnamefont {J.~S.}\ \bibnamefont
  {Al-Khalili}}\ and\ \bibinfo {author} {\bibfnamefont {J.~A.}\ \bibnamefont
  {Tostevin}},\ }\bibfield  {title} {\bibinfo {title} {{Matter Radii of Light
  Halo Nuclei}},\ }\href {https://doi.org/10.1103/PhysRevLett.76.3903}
  {\bibfield  {journal} {\bibinfo  {journal} {Phys. Rev. Lett.}\ }\textbf
  {\bibinfo {volume} {76}},\ \bibinfo {pages} {3903} (\bibinfo {year}
  {1996})}\BibitemShut {NoStop}%
\bibitem [{\citenamefont {N\"ortersh\"auser}\ \emph {et~al.}(2009)\citenamefont
  {N\"ortersh\"auser} \emph {et~al.}}]{PhysRevLett.102.062503}%
  \BibitemOpen
  \bibfield  {author} {\bibinfo {author} {\bibfnamefont {W.}~\bibnamefont
  {N\"ortersh\"auser}} \emph {et~al.},\ }\bibfield  {title} {\bibinfo {title}
  {{Nuclear Charge Radii of $^{7,9,10}\mathrm{Be}$ and the One-Neutron Halo
  Nucleus $^{11}\mathrm{Be}$}},\ }\href
  {https://doi.org/10.1103/PhysRevLett.102.062503} {\bibfield  {journal}
  {\bibinfo  {journal} {Phys. Rev. Lett.}\ }\textbf {\bibinfo {volume} {102}},\
  \bibinfo {pages} {062503} (\bibinfo {year} {2009})}\BibitemShut {NoStop}%
\bibitem [{\citenamefont {Wang}\ \emph {et~al.}(2021)\citenamefont {Wang},
  \citenamefont {Huang}, \citenamefont {Kondev}, \citenamefont {Audi},\ and\
  \citenamefont {Naimi}}]{Wang_2021}%
  \BibitemOpen
  \bibfield  {author} {\bibinfo {author} {\bibfnamefont {M.}~\bibnamefont
  {Wang}}, \bibinfo {author} {\bibfnamefont {W.}~\bibnamefont {Huang}},
  \bibinfo {author} {\bibfnamefont {F.}~\bibnamefont {Kondev}}, \bibinfo
  {author} {\bibfnamefont {G.}~\bibnamefont {Audi}},\ and\ \bibinfo {author}
  {\bibfnamefont {S.}~\bibnamefont {Naimi}},\ }\bibfield  {title} {\bibinfo
  {title} {{The AME 2020 atomic mass evaluation (II). Tables, graphs and
  references*}},\ }\href {https://doi.org/10.1088/1674-1137/abddaf} {\bibfield
  {journal} {\bibinfo  {journal} {Chin. Phys. C}\ }\textbf {\bibinfo {volume}
  {45}},\ \bibinfo {pages} {030003} (\bibinfo {year} {2021})}\BibitemShut
  {NoStop}%
\bibitem [{\citenamefont {Summers}\ and\ \citenamefont
  {Nunes}(2008)}]{PhysRevC.78.011601}%
  \BibitemOpen
  \bibfield  {author} {\bibinfo {author} {\bibfnamefont {N.~C.}\ \bibnamefont
  {Summers}}\ and\ \bibinfo {author} {\bibfnamefont {F.~M.}\ \bibnamefont
  {Nunes}},\ }\bibfield  {title} {\bibinfo {title} {{Extracting
  ($n,\ensuremath{\gamma}$) direct capture cross sections from Coulomb
  dissociation: Application to
  $^{14}\mathrm{C}$($n,\ensuremath{\gamma}$)$^{15}\mathrm{C}$}},\ }\href
  {https://doi.org/10.1103/PhysRevC.78.011601} {\bibfield  {journal} {\bibinfo
  {journal} {Phys. Rev. C}\ }\textbf {\bibinfo {volume} {78}},\ \bibinfo
  {pages} {011601(R)} (\bibinfo {year} {2008})}\BibitemShut {NoStop}%
\bibitem [{\citenamefont {Schmitt}\ \emph {et~al.}(2012)\citenamefont {Schmitt}
  \emph {et~al.}}]{PhysRevLett.108.192701}%
  \BibitemOpen
  \bibfield  {author} {\bibinfo {author} {\bibfnamefont {K.~T.}\ \bibnamefont
  {Schmitt}} \emph {et~al.},\ }\bibfield  {title} {\bibinfo {title} {{Halo
  Nucleus $^{11}\mathrm{Be}$: A Spectroscopic Study via Neutron Transfer}},\
  }\href {https://doi.org/10.1103/PhysRevLett.108.192701} {\bibfield  {journal}
  {\bibinfo  {journal} {Phys. Rev. Lett.}\ }\textbf {\bibinfo {volume} {108}},\
  \bibinfo {pages} {192701} (\bibinfo {year} {2012})}\BibitemShut {NoStop}%
\end{thebibliography}%
\end{document}